\documentclass[fleqn, usenatbib]{mnras}

\usepackage{newtxtext,newtxmath}
\usepackage[T1]{fontenc}

\usepackage{graphicx}
\usepackage{amsmath}
\usepackage{mathtools}
\usepackage{xcolor}
\usepackage{booktabs} 
\usepackage{textcomp} 
\usepackage{fix-cm}
\usepackage{float}
\usepackage{multirow}
\usepackage{siunitx}
\usepackage{csquotes}
\usepackage[capitalise]{cleveref}

\DeclareSIUnit{\pc}{pc}
\DeclareSIUnit{\kpc}{\kilo\pc}
\DeclareSIUnit{\Mpc}{\mega\pc}
\DeclareSIUnit{\yr}{yr}
\DeclareSIUnit{\Myr}{\mega\yr}
\DeclareSIUnit{\Gyr}{\giga\yr}
\DeclareSIUnit{\Msun}{\ensuremath{\mathrm{M}_\odot}}

\newcommand{\pdiff}[2]{\frac{\partial #1}{\partial #2}}

\newcommand{\brac}[1]{\left(#1 \right)}
\newcommand{\sbrac}[1]{\left[ #1 \right]}
\newcommand{\abrac}[1]{\left\langle #1 \right\rangle}

\newcommand{\bfrac}[2]{\left(\frac{#1}{#2} \right)}
\newcommand{\qmarks}{\enquote}
\newcommand{\eqn}[1]{\begin{equation} #1 \end{equation}}
\newcommand{\fft}[1]{\mathrm{fft} \sbrac{#1}}
\newcommand{\ifft}[1]{\mathrm{ifft} \sbrac{#1}}

\title[FDM impacts on primordial gas]{The Impact of Fuzzy Dark Matter Dynamics on the Accumulation and Fragmentation of Primordial Gas}

\author[A.\ F.\ Tocher et al.]{
Alexander Tocher,$^{1}$\thanks{E-mail: at802@cam.ac.uk}
Anastasia Fialkov,$^{1}$
Simon May,$^{2}$
Ralf S.\ Klessen,$^{3,4}$
Simon C.~O.\ Glover,$^{3}$
Paul C.\ Clark,$^{5}$ \newauthor
and Tibor Dome$^{1}$
\\
$^{1}$Institute of Astronomy, Madingley Road, Cambridge, CB3 0HA, UK\\
$^{2}$Fakultät für Physik, Universität Bielefeld, Universitätsstraße 25, 33615 Bielefeld, Germany
\\
$^{3}$Universität Heidelberg, Zentrum für Astronomie, Institut für Theoretische Astrophysik, Albert-Ueberle-Str. 2, D-69120 Heidelberg, Germany\\
$^{4}$Universität Heidelberg, Interdisziplinäres Zentrum für Wissenschaftliches Rechnen, Im Neuenheimer Feld 205, D-69120 Heidelberg, Germany\\
$^{5}$School of Physics and Astronomy, Queen's Buildings, The Parade, Cardiff University, Cardiff, CF24 3AA, UK\\
}

\date{Accepted XXX. Received YYY; in original form ZZZ}

\pubyear{2026}

\begin{document}

\label{firstpage}
\pagerange{\pageref{firstpage}--\pageref{lastpage}}
\maketitle


\begin{abstract}
{%
Fuzzy Dark Matter (FDM), particularly in the $10^{-22}$\,eV mass regime is frequently used to characterize wave-like interference effects. It exhibits macroscopic wave properties, which drive distinct baryonic dynamics within collapsed haloes. Using the hydrodynamical code \textsc{arepo} with the \textsc{axirepo} module and primordial chemistry, we simulate the assembly of haloes with masses $\SI{3e8}{\Msun} \le M_{\mathrm{h}} \le \SI{8e9}{\Msun}$ across a range of axion masses $\SI{1e-22}{\eV} \le m_{\mathrm{a}} \le \SI{7e-22}{\eV}$. We investigate how small-scale dynamics of the FDM density field affect the accumulation of cold, dense gas essential for primordial star formation. We demonstrate that gas collapse is suppressed by a two-fold mechanism: a delay driven by the geometry of the FDM solitonic core and a secondary dynamical barrier caused by stochastic wave fluctuations. While the flattened solitonic potential profile itself inhibits central gas accumulation, these wave-driven dynamics provide a further layer of disruption and angular momentum support, which in certain regimes prevents gas from reaching the central, compact, high-density configurations characteristic of CDM. Consequently, sites of star formation are shifted away from a single central peak toward a population of lower-mass clusters. Our work provides a physical framework for calibrating halo mass-dependent star formation efficiencies in FDM cosmologies, where internal processes may delay Cosmic Dawn beyond the effects of the initial power spectrum cut-off. These results are essential for interpreting realistic observational constraints from future 21-cm signal observations and the faint-end luminosity functions observed by the JWST, as well as providing an upper bound on the baryonic effects in the context of Mixed Dark Matter scenarios.}

\end{abstract}
  

%%%%%%%%%%%%%%%%%%%%%%%%%%%%%%%%%%%%%%%%%%%%%%%%%%

%%%%%%%%%%%%%%%%% BODY OF PAPER %%%%%%%%%%%%%%%%%%

\section{Introduction}
\label{sec:introduction}

Among the candidates for Dark Matter (DM), the standard Cold Dark Matter (CDM) paradigm has been remarkably successful in explaining the large-scale structure of the Universe \citep{Springel2005}. Historically, Fuzzy Dark Matter (FDM) {with an axion-like particle mass of $m_{\mathrm{a}} \sim \SI{e-22}{\eV}$} was proposed as an alternative to CDM that maintained these large-scale successes while potentially resolving persistent challenges observed on sub-galactic scales \citep{Boylan2011, Weinberg2015, DelPopolo2017}. These discrepancies, often referred to as the \qmarks{cusp–core} problem, the \qmarks{too big to fail} problem, and the \qmarks{missing satellites} problem \citep{Weinberg2015, Marsh2016, Bullock2017, Hui2017, Eberhardt2025}, involve an over-prediction of the density and abundance of small-scale structures in DM-only simulations. {The status of these problems has shifted in recent years, with many of these issues now largely considered to be resolvable within the CDM framework by accounting for baryonic feedback processes. For example,} supernova-driven gas outflows and reionization can flatten central density profiles and suppress the formation of low-mass satellites \citep[e.\,g.,][]{Wetzel2016, GarrisonKimmel2019, Sales2022}. {Moreover, models in which FDM comprises \SI{100}{\percent} of DM have been significantly constrained by modern observations. High-resolution probes of the Lyman-$\alpha$ forest currently disfavour monolithic models with $m_{\mathrm{a}} \lesssim \SI{2e-20}{\eV}$ \citep{Irsic2017, Armengaud2017, Kobayashi2017, Rogers2021}, while studies of dynamical heating in ultra-faint dwarfs (UFDs) have pushed these lower bounds as high as $m_{\mathrm{a}} \gtrsim \SI{3e-19}{\eV}$, or even $m_{\mathrm{a}} \gtrsim \SI{8e-18}{\eV}$ if objects like Ursa Major~III/UNIONS~1 are confirmed as being galaxies \citep{Dalal2022, May2025}.}

{%
Despite this conceptual shift and the mounting observational evidence, it is premature to fully exclude the $m_\mathrm{a} \sim \SI{e-22}{\eV}$ regime as a physical possibility, because the observational landscape for dwarf galaxies remains complex. While some systems suggest high $m_a$ bounds, other recent analyses of dark matter-dominated dwarf populations continue to find evidence favouring $m_\mathrm{a} \sim \SI{e-22}{\eV}$ mass range \citep{Indjin2026, Korshynska2026}. This suggests that the observational signatures of FDM may be highly sensitive to the specific evolutionary history of individual haloes. Moreover, FDM with $m_\mathrm{a} \sim \SI{e-22}{\eV}$ remains a highly appealing and widely studied dark matter scenario owing to its unique, testable predictions arising from fundamental physics \citep{Hui2017, Chowdhury2023, May2023, Mocz2023, Liu2023, Shen2024, Liao2025, Eberhardt2025, Alvarez2025, Nadler2025} as well as the limiting case of Mixed Dark Matter (MDM) models in which FDM comprises a fraction of dark matter \citep{Schwabe2020, Lague2024, Dome2025, Wang2026}. Motivated by these considerations, we simulate monolithic FDM in the $m_{\mathrm{a}} = \SIrange{1e-22}{7e-22}{\eV}$ range to explore the interplay between wave-driven dynamics and primordial collapsing gas. Further motivation is provided in Section \ref{sec:motivation} of this paper.}

{In the classical monolithic} FDM {scenario, the DM} consists of ultra-light bosonic particles, often identified as ultra-light axions, with masses on the order of \SI{e-22}{\eV} \citep{Hu2000, Marsh2014, Ferreira2021, Chavanis2025}. These particles are similar to those derived from the axion theory initially developed to solve the strong CP problem in quantum chromodynamics. The extremely low mass of FDM particles results in macroscopic wave effects, where their de Broglie wavelength is on $\sim \si{\kpc}$ scales, comparable to the sizes of dwarf galaxies. The wavelike nature of FDM introduces novel dynamics into the behaviour of dark matter on astrophysical scales: the very large de Broglie scale provides  \qmarks{quantum pressure} that naturally smooths central cusps and suppresses the formation of (sub-)haloes below a certain mass threshold.

On cosmological scales, the suppression of small-scale structures in FDM models results in fewer low-mass dark matter haloes forming compared to predictions from CDM \citep[e.\,g.][]{Kulkarni2022, May2023}. This has direct implications for the cooling and condensation of gas within these haloes, which are the sites of early star formation. In the standard CDM paradigm, first star (Population III) formation occurs in \qmarks{minihaloes} of \SIrange{e5}{e6}{\Msun} where gas collapse is driven by molecular hydrogen ($\mathrm{H}_2$) cooling \citep{Bromm2004, Glover2013}. Because FDM suppresses these small-scale fluctuations, primordial star formation is shifted to more massive atomic cooling haloes ($M_{\mathrm{h}} \gtrsim \SIrange{e7}{e9}{\Msun}$), where the virial temperature reaches \SI{10000}{\K}. At these temperatures, the initially hot gas can cool down via efficient Ly$\alpha$ emission, initiating the runaway collapse phase ($n > \SI{e2}{\per\cm\cubed}$). In this work, we employ \qmarks{sink particles} as a numerical proxy for this cold dense gas, representing the pre-stellar collapse phase and allowing us to track the star-formation potential of each halo.

To date, there have been many studies that investigate the large scale properties of FDM cosmologies, both with and without gas \citep[e.\,g.][]{Woo2009, Schive2014, Veltmaat2018, Mocz2017, Mocz2019, Mocz2020, May2021, May2023, Dome2023b, Dome2023, Dome2024, Dome2025}. However, few studies have captured the smaller-scale effects of the FDM dynamics on the collapsing gas and subsequent star formation, which we consider in this paper.  While earlier works such as \citet{Hirano2018} looked at the global shift in the timing and sites of first star formation in FDM cosmologies, and \citet{Kulkarni2022} or \citet{Yang2024} considered how modified potential profiles and interference patterns might inhibit gas condensation and stellar density profiles, our innovation lies in capturing the full interplay between the wave-interference dynamics, the primordial chemistry of the gas, and the gas physics of collapse. The wave-like nature of FDM, including phenomena like soliton formation and interference patterns, leads to granular density fluctuations in the dark matter on small scales \citep{Hu2000}. We might expect that these persistent dynamics would result in induced turbulence within the collapsing gas, which has been shown in the context of CDM simulations \citep[e.\,g.][]{Klessen2010, Stacy2011, Krumholz2018, Soam2024} to be a key factor in star formation processes.

The gravitational effects of these fluctuations on stellar dynamics have been studied through simulations of local structures \citep{Amorisco2018, Church2019, Marsh2019, Dalal2022, May2025}. These studies demonstrate that collapsed FDM structures can induce observable effects on embedded baryons. However, to date, there has been little work on how these dynamics affect the collapse phase in the earliest haloes.

Calibrating these star formation models is essential for {developing a crucial bookend case for the more complex MDM models as well as producing} more reliable predictions for high-redshift galaxy observations with JWST and 21-cm experiments like the Square Kilometer Array (SKA). While JWST has made revolutionary measurements of high-redshift galaxies, particularly at the bright and massive end of the UV luminosity function (UVLF) \citep{Finkelstein2023}, observations from surveys like the Hubble Frontier Fields are providing constraints on the faint end of the UVLF, which are most sensitive to FDM particle masses \citep{Sipple2025}. A delay or disruption in early star formation could also have observable consequences for the 21-cm signal, potentially placing lower limits on the FDM particle mass \citep{Lidz2018, Nebrin2019, Flitter2022}. Importantly, these 21-cm studies typically only account for the suppression of the halo mass function, thus ignoring the additional suppression in star formation introduced by small-scale FDM dynamics; we aim to improve on this by investigating how the intrinsic FDM dynamics further impact the small-scale star formation efficiency. 

This work aims to bridge the existing gap in the literature by considering the impact of FDM dynamics on primordial star-forming gas across a range of axion-like particle masses ($m_{\mathrm{a}} = \SIrange{1e-22}{7e-22}{\eV}$) and DM halo masses ($M_{\mathrm{h}} = \SIrange{3e8}{8e9}{\Msun}$). We create numerical simulations of FDM using the \textsc{axirepo} code within the framework of \textsc{arepo} \citep{Springel2010, May2021, May2023} with the addition of a primordial chemistry network \citep{Clark2011}, which we use to model both the atomic and molecular hydrogen ($\mathrm{H}_2$) cooling necessary to track first star formation in FDM environments. A summary of our simulations can be found in \cref{tab:simulation_parameters}. This paper is organized as follows:  in \cref{sec:motivation} we discuss the continuous motivation for FDM, in \cref{sec:theory} we outline the FDM theory, followed by methodology in \cref{sec:methods}. We then present our results in \cref{sec:results} and discuss the implications in \cref{sec:discussion}. A summary of our conclusions can be found in \cref{sec:conclusion}.

\begin{table*}
    \centering
    \caption{Summary of simulations run. FDM runs are characterized by a fixed spatial grid resolution $\Delta x = \SI{31.25}{\pc}$ (in physical \si{\pc}), while CDM runs are characterized by a mass resolution of \SIlist{480; 180; 48; 18}{\Msun} for our largest to smallest haloes (following the first row of the table). The dashes \qmarks{---} represent cases that were omitted (e.\,g., due to tidal instability or being outside the relevant parameter space).}
    \label{tab:simulation_parameters}
    \begin{tabular}{c cc cc cc cc c}
    \toprule
     & \multicolumn{8}{c}{\textbf{FDM models ($m_{\mathrm{a}}$) }} & \textbf{CDM} \\
      & \multicolumn{2}{c}{\SI{1e-22}{\eV}} & \multicolumn{2}{c}{\SI{2e-22}{\eV}} & \multicolumn{2}{c}{\SI{3e-22}{\eV}} & \multicolumn{2}{c}{\SI{7e-22}{\eV}} & \\
    \textbf{Halo Mass} (\si{\Msun}) & Dynamic & Frozen & Dynamic & Frozen & Dynamic & Frozen & Dynamic & Frozen &  \\
    \midrule
    $8 \times 10^9$ & \checkmark &  \checkmark &  --- & --- & \checkmark &  \checkmark &  \checkmark &  \checkmark &  \checkmark\\
    $3 \times 10^9$ & \checkmark &  \checkmark &  \checkmark &  \checkmark &  \checkmark &  \checkmark &  \checkmark &  \checkmark &  \checkmark\\
    $8 \times 10^8$ & --- & --- & \checkmark &  \checkmark &  \checkmark &  \checkmark &  \checkmark &  \checkmark &  \checkmark\\
    $3 \times 10^8$ & --- & --- & --- & --- & \checkmark &  \checkmark &  \checkmark &  \checkmark &  \checkmark\\
    \bottomrule
    \end{tabular}
\end{table*}

\section{Motivation}
\label{sec:motivation}

FDM remains a highly appealing dark matter scenario which yields numerous unique, testable predictions arising from fundamental physics \citep[e.\,g.][]{Eberhardt2025} and providing the boundary case for the increasingly prominent MDM models \citep[e.\,g.][]{Lague2024, Dome2025}.

Current dwarf galaxy observations present a diverse and sometimes contradictory picture of the axion mass. While many frameworks rule out $m_{\mathrm{a}} \sim \SI{1e-22}{\eV}$ as a monolithic candidate, the continued emergence of signatures favouring this scale \citep{Indjin2026, Korshynska2026} points toward a more nuanced reality. It is likely that the resulting observational footprints are sensitive to the specific evolutionary pathways of individual structures, rather than being a simple function of the global dark matter model.

Furthermore, and most critical for this work, the \SI{e-22}{\eV} mass range serves as a fundamental benchmark for more complex, viable cosmologies such as MDM or \qmarks{axiverse} scenarios \citep{Arvanitiki2010, Gosenca2023, Lague2024, Eberhardt2025, Dome2025}. From a particle physics perspective, the presence of only a single ultra-light dark matter species, although most commonly used in studies of structure formation for reasons of simplicity and efficiency, would indeed be rather unlikely. By relaxing this assumption of a monolithic dark matter sector, MDM models provide a physically motivated framework to potentially reconcile the strongest large-scale structure constraints from the Lyman-$\alpha$ forest \citep{Rogers2021} and halo mass functions \citep{Sipple2025} with the stringent dynamical bounds derived from local dwarf galaxies \citep{Dalal2022, May2025}. As demonstrated by \citet{Dome2025}, MDM haloes exhibit a halo mass-dependent axion depletion effect where low-mass haloes (precisely those used to derive the most stringent bounds $> \SI{e-19}{\eV}$) are significantly depleted of their axion component relative to the cosmic mean. These systems are effectively CDM-like in their internal structure, significantly suppressing the efficiency of any dynamical heating relative to the monolithic case, while still allowing for a significant $m_\mathrm{a} \sim \SI{e-22}{\eV}$ component to exist globally.%

Studies of classical monolithic FDM in the $m_{\mathrm{a}} = \SIrange{1e-22}{7e-22}{\eV}$ range are further motivated by scalability. The FDM dynamics are governed by the Schrödinger--Poisson equations, which admit a well-known scaling relation \citep[see e.\,g.][]{Mocz2017}. Under this scaling, the dynamics of a well-understood halo can be mapped onto a new system with a different halo mass and $m_a$. While it is important to note that baryonic physics (such as gas cooling or stellar feedback) do not scale in this way and require independent investigation, the gravitational behaviour of the FDM remains qualitatively similar across scales. Therefore, resolving the regime where the dynamics are most prominent is a necessary prerequisite to understanding the effects that FDM dynamics can have on the primordial gas. By isolating these mechanisms at computationally resolvable systems such as explored in this work, we can establish a pathway towards for the broader class of wave-dark matter models at different $m_\mathrm{a}$, regardless of whether the axion is a monolithic or sub-dominant dark matter component.%

Considering this strong continuous motivation, we simulate \qmarks{classical monolithic} FDM models in the $m_{\mathrm{a}} = \SIrange{1e-22}{7e-22}{\eV}$ range in a controlled and computationally feasible setup to characterize the interactions between the FDM dynamics and the baryonic sector. We explore the implications of the wave-driven dynamics on the collapse and distribution of cold, potentially star-forming, gas. With this work, we aim to provide the community with a crucial bookend case for the more complex MDM models and to enable new independent FDM/MDM constraints at Cosmic Dawn.%

\section{Theoretical Background}
\label{sec:theory}

For this work, we consider a physical system composed of two primary components: the dark matter, which provides the underlying gravitational potential and cosmic scaffolding, and the primordial gas, which accretes into these potential wells. The physical problem centres on how the unique small-scale dynamics of FDM modify the standard process of gas cooling and collapse. While the dark matter serves as the gravitational driver, its wave-like nature in the FDM model creates a non-static potential, characterized by a central solitonic core and persistent interference-induced fluctuations. Our setup investigates the \qmarks{active} role of these FDM dynamics, focusing on how they exert time-varying gravitational influences that can hinder gas accretion and alter the timing of the first star formation events in the early universe.

In the following Section, we lay out the fundamental theory underlying the FDM model (\cref{sec:theory:FDM}), and the important dynamics within FDM haloes (\cref{sec:theory:dynamics}), both of which are key in understanding the resulting effects on gas collapse.

\subsection{Ultralight Dark Matter}
\label{sec:theory:FDM}

In the non-relativistic limit and at zero temperature, a spin-0 ultra-light scalar field represented by the (collective, mean-field) \qmarks{wavefunction} $\psi$ is governed by the Schrödinger–Poisson (SP) equations \citep{Hu2000}:%
{\allowdisplaybreaks
\begin{align}
    i\hbar \pdiff{\psi}{t} &= - \frac{\hbar^2}{2m_{\mathrm{a}}} \nabla^2 \psi + m_{\mathrm{a}} V \psi,
    \label{eqn:Schrodinger}
    \\
    \nabla^2 V &= 4 \pi G \brac{\rho - \bar{\rho}}.\label{eqn:Poisson}
\end{align}%
}
In the above equations, $\psi$ is the FDM \qmarks{wavefunction} normalized such that the dark matter density $\rho_{\mathrm{dm}} = |\psi|^2$, $m_{\mathrm{a}}$ is the boson mass, and $V$ is the gravitational potential which is affected by FDM as well as the baryonic component via the Poisson \cref{eqn:Poisson}, where the total matter density is given by $\rho = \rho_{\mathrm{dm}} + \rho_{\mathrm{gas}}$ and $\bar{\rho}$ is the mean matter density.

Instead of the wavefunction, the axion-like field can be represented by a fluid with density and velocity, related to the complex wavefunction via the Madelung transformation \citep{Madelung1927, Suarez2011, Chavanis2011}, where we decompose the $\psi$ into an amplitude $\sqrt{\rho_{\mathrm{dm}}}$ and a phase $S$. Given this decomposition we now have
\eqn{
    \begin{aligned}
    \psi &= \sqrt{\rho_{\mathrm{dm}}} e^{iS/\hbar} ,\\  %\qquad, \qquad
    v &= \nabla S /m_{\mathrm{a}}.
    \end{aligned}
    \label{eqn:madelung}
}

The SP equations obey both conservation of mass:
\begin{align}
    M &= \int \rho_{\mathrm{dm}} \, \mathrm{d}^3x,
    \\
\shortintertext{and energy:}
    E_\mathrm{tot} &= \int \sbrac{  \frac{\hbar^2}{2m^2} |\nabla \psi|^2 + \frac{1}{2} V |\psi|^2  } \mathrm{d}^3x  \label{eqn:energy}\\
    % &= \int\frac{\hbar^2}{2m^2} \sbrac{\Re(\nabla \psi)^2 + \Im( \nabla \psi)^2} d^3x + \int \frac{1}{2} \rho_{\mathrm{dm}}  V  d^3x \\
    &= \int\frac{\hbar^2}{2m^2} \brac{\nabla \sqrt{\rho_{\mathrm{dm}}}}^2 \, \mathrm{d}^3x  + \int\frac{1}{2} \rho_{\mathrm{dm}} v^2 \, \mathrm{d}^3x + \int \frac{1}{2} \rho_{\mathrm{dm}}  V  \, \mathrm{d}^3x \notag \\
    &= K_\rho + K_v + W , \notag
\end{align}
where the integration volume is any closed volume. The total kinetic energy $K = K_\rho + K_v$ is given by the sum of  the energy due to the quantum pressure tensor $K_\rho$ and  the classical contribution $K_v$. $W$ is the potential energy. The decomposition of the kinetic energy term is obtained by substituting the Madelung form into the gradient $|\nabla \psi|^2$. The phase information $S$ is mapped directly onto the bulk velocity term $K_v$ via its gradient ($\nabla S$), while the absolute value of the wavefunction amplitude yields the quantum pressure term $K_\rho$. Note that in the potential energy term $W$, the phase factor $e^{iS/\hbar}$ disappears entirely when taking the inner product $|\psi|^2 = \psi \psi^*$, leaving only the dependence on the density $\rho_{\mathrm{dm}}$.

For $m_{\mathrm{a}}$ below approximately $\SI{e-18}{\eV}$, the FDM density field exhibits wave behaviour on astrophysical scales. The characteristic scale of these fluctuations is the de Broglie wavelength, $\lambda_{\mathrm{dB}} = h/(m_{\mathrm{a}} \sigma)$. Here, $\sigma$ represents the effective velocity dispersion of the system. It is important to note that because the Madelung velocity $v$ (\cref{eqn:madelung}) is derived from a single-valued phase $S$, the local velocity dispersion is zero by definition. In this context, $\sigma$ is understood as a macroscopic quantity representing the root-mean-square velocity of the wave-field fluctuations, or equivalently, as a measure of the total kinetic energy within a given volume. For a typical velocity dispersion of $\sigma = \SI{100}{\km\per\s}$ and $m_{\mathrm{a}} = \SI{e-22}{\eV}$, this yields a de Broglie scale (in physical units) of:
\eqn{\lambda_{\mathrm{dB}} \approx 4 \bfrac{\SI{e-22}{\eV}}{m_{\mathrm{a}}} \bfrac{\SI{100}{\km\per\s}}{\sigma} \,\si{\kpc}.  }

This wave-like behaviour results in distinctive observational signatures, most notably the suppression of small-scale structure formation \citep{Hu2000, Hui2017}. Within the centres of FDM haloes, quantum pressure counteracts gravitational collapse and introduces a \qmarks{quantum Jeans scale} \citep{Khlopov1985} which leads to the formation of solitonic cores \citep{Schive2014, Mocz2017}. These cores represent stable, high-density central regions where the density profile is maintained by the balance between gravity and quantum pressure, and can be approximated as:
\eqn{ \rho_\mathrm{dm}(r) = \frac{\rho_{\mathrm{c}}}{ \brac{1+\brac{\dfrac{r}{r_{\mathrm{c}}}}^2}^8}, }    
where $\rho_c$ is the central density and $r_{\mathrm{c}}$ is the core radius. This core and its properties dominate the central potential of an FDM halo and, therefore, regulate the collapse of the gas in its central regions.

\subsection{Dynamics of FDM Haloes}
\label{sec:theory:dynamics}

One of the unique signatures of the FDM physics is that the wavefunction interference produces ubiquitous oscillations in the local density field across the halo. These oscillations have an amplitude of order unity $\delta \rho_{\mathrm{dm}} \sim \rho_{\mathrm{dm}}$ and characteristic length and time scales of $r \approx \lambda_{\mathrm{dB}}/2\pi = \hbar /m_{\mathrm{a}} \sigma$ and $\tau_\mathrm{osc} \approx \lambda_{\mathrm{dB}}/v = h/m_{\mathrm{a}}\sigma^2$ where $\lambda_\mathrm{dB}$ is as defined above in \cref{sec:theory:FDM}. For a typical combination of $m_{\mathrm{a}} = \SI{e-22}{\eV}$ and $\sigma = \SI{100}{\km\per\s}$ from our work, this results in:
\begin{align}
r &\approx 0.6 \bfrac{\SI{e-22}{\eV}}{m_{\mathrm{a}}} \bfrac{\SI{100}{\km\per\s}}{\sigma} \,\si{\kpc} ,
\\
\tau_{\mathrm{osc}} &\approx 40 \bfrac{\SI{e-22}{\eV}}{m_{\mathrm{a}}} \bfrac{\SI{100}{\km\per\s}}{\sigma}^2 \,\si{\Myr} .
\end{align}
This phenomenon results in corresponding fluctuations in the gravitational potential within the halo \citep{Church2019, Marsh2019, Rios2022}.

The soliton at the centre of each FDM halo is also not static, but instead exhibits similar order-unity fluctuations in its central density (with a period of $\sim \tau_\mathrm{osc}$) as well as a random walk behaviour about the centre of mass of the halo with magnitude $\sim \lambda_{\mathrm{dB}}$ \citep{Schive2014, Schive2020, Hui2021}. These behaviours are a result of persistent interference of the wavefunction outside the soliton with the soliton itself. This is important as the soliton dominates the gravitational potential at small distances from the halo centre. We expect these effects are most pronounced in small haloes in models with a low $m_{\mathrm{a}}$, since the mass contained within the soliton scales roughly as $M_{\mathrm{s}} \propto M_{\mathrm{h}}^{1/3}m_{\mathrm{a}}^{-1}$ \citep{Schive2014b} \citep[although this relation shows significant variation, see e.\,g.][]{Chan2022} and thus the proportion of halo mass contained in the soliton can approximately be estimated as $M_{\mathrm{s}}/M_{\mathrm{h}} \propto M_{\mathrm{h}}^{-2/3}m_{\mathrm{a}}^{-1}$.

To illustrate these theoretical principles, we first examine a simplified dark matter-only (DMO) configuration. This allows us to isolate the dynamical behaviour of dark matter and confirm the theory results we have discussed before introducing the complexities of gas physics. In this simplified setup, the gravitational potential is determined solely by the FDM density, allowing us to isolate the wave-driven fluctuations of the core without the baryons which will be discussed in \cref{sec:results}. For this illustrative case, we consider an isolated halo of mass $M_{\mathrm{h}} \approx \SI{3e9}{\Msun}$ with an axion mass $m_{\mathrm{a}} = \SI{e-22}{\eV}$ within a \SI{32}{\kpc} cubic box and a dark matter resolution of $1/32\,\si{\kpc} = \SI{31.25}{\pc}$.

\begin{figure}
    \centering
    \includegraphics[width=0.95\linewidth]{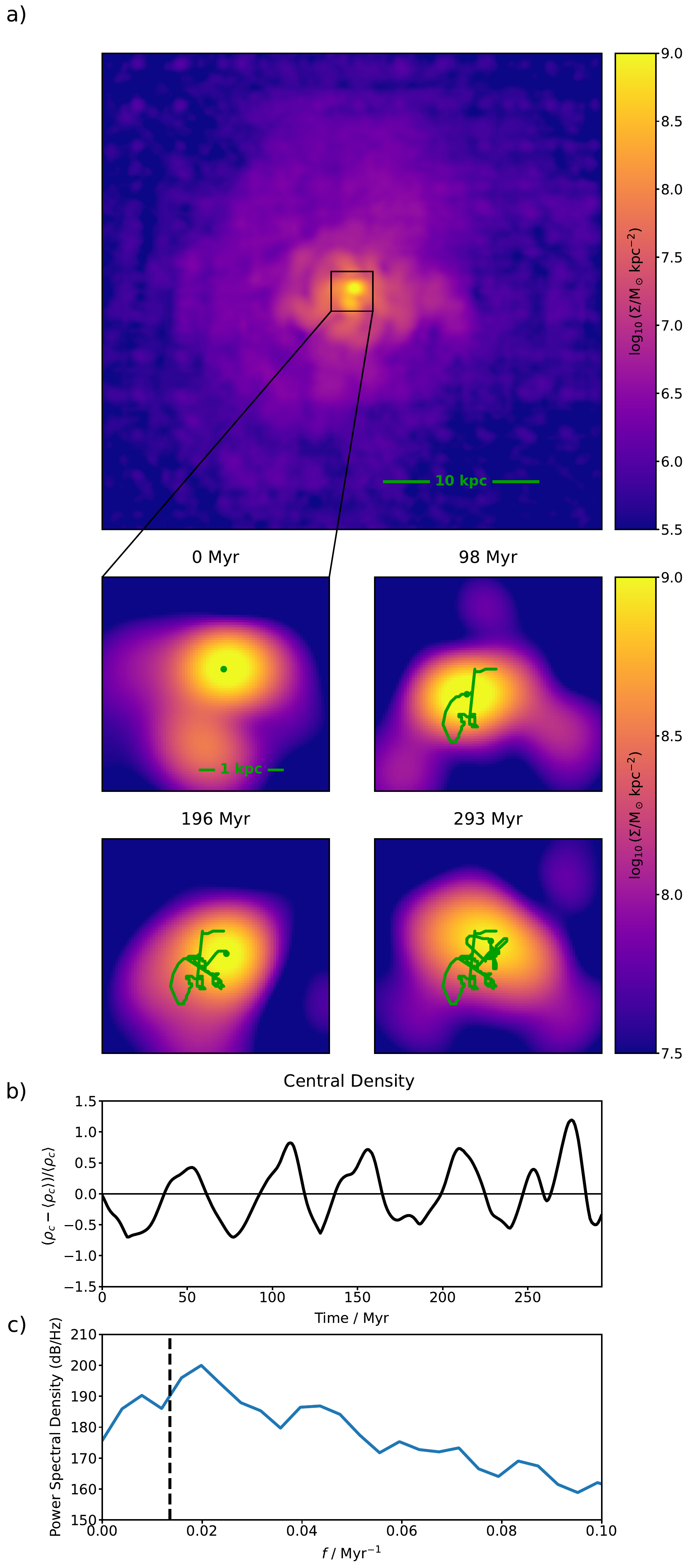}
    \caption{Dynamics of the solitonic core at the centre of a simulated FDM halo with a halo mass $M_{\mathrm{h}} = \SI{3e9}{\Msun}$ for axion mass $m_{\mathrm{a}} = \SI{e-22}{\eV}$. The panels marked a) show the projected (log) density of the dark matter $\rho_{\mathrm{dm}}$ of a zoomed-in region at the centre of an FDM halo for a series of snapshots separated by $\sim \SI{100}{\Myr}$. The green line shows a track of the central density from its initial position at 0 Myr within the halo sampled with time resolution $\Delta t \approx \SI{1}{\Myr}$ at four snapshots throughout the simulation, spaced approximately \SI{100}{\Myr} apart. The panel marked b) shows the order unity fluctuations of the central density along this track over time, with a characteristic period of $f \sim \SI{0.02}{\per\Myr} \Rightarrow \tau_\mathrm{osc} \sim \SI{50}{\Myr}$, which can be identified by the peak in the power spectral density of this oscillation in the bottom panel marked c). This lines up with the theoretical value for the period of oscillation $\tau_{\mathrm{osc}} \approx \SI{65}{\Myr}$ at the centre of the halo, shown by the dashed line.}
    \label{fig:soliton_evolution}
\end{figure}

The evolution of the soliton can clearly be seen in these simulations. \Cref{fig:soliton_evolution}~a) shows a series of snapshots from a dark matter-only halo zoomed in on the central region, where the green line shows the movement of the centre of the solitonic core sampled approximately every \SI{1}{\Myr} over the course of four snapshots separated by approximately \SI{100}{\Myr}. Each zoomed panel shows the same simulation box coordinates for consistency between the panels. The panel marked b) shows how the central density $\rho_{\mathrm{c}}$, defined as the density of the densest point of the DM, evolves over time, with order unity fluctuations in the central density contrast $\brac{\rho_c - \abrac{\rho_{\mathrm{c}}}}/\abrac{\rho_{\mathrm{c}}}$ clearly visible, where $\abrac{\rho_{\mathrm{c}}}$ is the time averaged central density of the halo. The oscillation timescale $\tau_\mathrm{osc} $ closely matches the theory value (shown as the dashed vertical line) and for this halo with $m_{\mathrm{a}} = \SI{1e-22}{\eV}$ and $\sigma \approx \SI{80}{\km\per\s}$ corresponds to $\tau_{\mathrm{osc}} \approx \SI{50}{\Myr}$.

In this work, the centre of each dark matter halo is defined as the minimum of its gravitational potential unless otherwise specified. This approach is adopted in preference to its the centre of mass because it is a more physically relevant quantity that governs the dynamics of the central gas. Since the potential fluctuations at the centre of the halo are the primary mechanism we hypothesize to affect gas collapse and star formation, using the centre of potential provides a physically motivated and direct reference frame for analysing the response of the gas to these small-scale dynamics. In FDM, this point undergoes a stochastic random walk due to soliton migration, as shown in \cref{fig:soliton_evolution}~a). All radial profiles and analyses presented in this paper are centred on this time-varying point unless explicitly stated otherwise.

\section{Methodology}
\label{sec:methods}

In this section, we describe the numerical framework and experimental design used to investigate gas collapse processes in the first FDM haloes. Our methodology is structured to isolate the physical effects of wave-driven fluctuations from standard gravitational collapse.

First, in \cref{sec:methods:design}, we outline our simulation design, including the three-way comparison between CDM, FDM, and \qmarks{frozen} FDM models. The computational details of the \textsc{arepo} and \textsc{axirepo} solvers, including the gravitational coupling between dark matter and baryons, are detailed in \cref{sec:methods:code}. Lastly, we describe the generation of our initial conditions and the physical motivation for our choice of halo and axion-like particle masses in \cref{sec:methods:ics}.

\subsection{Simulation Design and Objectives}
\label{sec:methods:design}

The primary goal of this work is to isolate and quantify how the unique small-scale dynamics of FDM, specifically the fluctuations of the central solitonic core, influence the collapse of primordial gas and the assembly of the dense star-forming regions. To achieve this, we compare the evolution of gas infall onto a an isolated dark matter halo across three distinct dark matter scenarios: standard CDM, fully simulated \qmarks{dynamic} FDM, and a \qmarks{frozen} FDM model (explained below). For the remainder of this work \qmarks{dynamic} FDM refers to those runs where the SP equations (\cref{eqn:Schrodinger,eqn:Poisson}) are solved and the FDM evolves over time.

For our simulation suite, summarized in \cref{tab:simulation_parameters}, we chose four halo masses at \SIlist{3e8; 8e8; 3e9; 8e9}{\Msun}. We then evolved each of these initial setups for four axion masses (\SIlist{e-22; 2e-22; 3e-22; 7e-22}{\eV}), as well as the CDM comparison case. All simulations are performed in fixed physical boxes with side length $L_{\mathrm{box}} = \SI{32}{\kpc}$ and with periodic boundary conditions which are necessary for the pseudo-spectral SP solver \citep[see \cref{sec:methods:code:FDM} and ][]{Mocz2017, May2021}. A box size of $L_{\mathrm{box}} = \SI{32}{\kpc}$ is sufficient for our needs; at redshifts $z \sim \numrange{10}{20}$, the virial radii for the haloes in our sample range from a few hundred pc to a few \si{\kpc} \citep{Bryan1998, Barkana2001}. Thus, our \SI{32}{\kpc} box means that the isolated system remains well-defined. This choice is also motivated by our need to resolve the internal dynamics of the central $\sim$ kiloparsec. The dark matter resolution is handled according to the specific solver:
\begin{enumerate}

    \item \textbf{FDM resolution:} All FDM runs (both \qmarks{dynamic} and \qmarks{frozen}) utilize a fixed $1024^3$ Cartesian grid, providing a uniform spatial resolution of $\Delta x_{\mathrm{dm}} = \SI{31.25}{\pc}$. Numerical convergence in FDM simulations requires that the solitonic core radius be resolved by at least four grid cells to accurately capture the dynamics of the soliton and central density fluctuations \citep{Mocz2017, May2021}.
    
    \item \textbf{CDM resolution:} Our CDM simulations use $256^3$ $N$-body particles. This corresponds to a mass resolution ($m_{\mathrm{DM}}$) ranging from $\sim \SI{18}{\Msun}$ for our \SI{3e8}{\Msun} halo to $\sim \SI{480}{\Msun}$ for our \SI{8e9}{\Msun} halo. To resolve the central density profile, we employ a gravitational softening length of $\epsilon = \SI{3.1}{\pc}$. This scale is an order of magnitude smaller than the fixed spatial grid resolution used in our FDM simulations, ensuring that the central cusp is accurately captured without numerical flattening. This disparity in resolution is intentional, as it allows the CDM cases to provide a true cusp at the halo centre, whereas the flattened FDM density profile does not require such high resolution.
    
    \item \textbf{Gas resolution:} The gas component is initialized with a total mass in line with the cosmological baryon fraction and using $256^3$ gas cells. Initially this corresponds to a gas cell mass of $m_{\mathrm{gas}} \approx \SI{3}{\Msun}$ for our lowest halo mass, and up to $m_{\mathrm{gas}} \approx \SI{90}{\Msun}$ for our highest halo mass. To accurately capture the fragmentation of primordial gas and avoid numerical artifacts, we employ a Jeans refinement criterion. We ensure that the local Jeans length is resolved by a minimum of 16 cell diameters throughout the simulation. This exceeds the standard Truelove criterion \citep{Truelove1997} by a factor of four, ensuring that the fragmentation observed in the solitonic cores is physically driven by the interplay of gravity, thermal pressure, and FDM-induced fluctuations.
    
\end{enumerate}

To ensure these results are numerically robust, we performed convergence tests on a $2048^3$ grid for the $M_{\mathrm{h}} = \SI{8e9}{\Msun}$ case. Even with the highest axion mass ($m_{\mathrm{a}} = \SI{7e-22}{\eV}$), which presents the smallest physical substructures to resolve, we saw $< 2\%$ deviation in key metrics used in this paper between the $1024^3$ and $2048^3$ grids. 

The choice to use a box of fixed physical size was motivated by two factors: 
\begin{enumerate}
    \item In a box of fixed physical size, our fixed-grid FDM spatial resolution is not degraded over time, as the universe expands.
    \item For studying individual haloes, once they have collapsed, they have become decoupled from the cosmological expansion and so the expansion should not affect the internal dynamics of the halo.
\end{enumerate}

The wave-mechanical nature of FDM imposes a strict constraint on the simulation time step; to evolve the phase of the scalar field correctly, the maximum allowed time step satisfies $\Delta t \propto \Delta x^2$ (the Schrödinger Courant condition), which is significantly more restrictive than the standard gravitational Courant condition used in our CDM runs \citep{Mocz2017, May2021}. Resolving these dynamics while simultaneously tracking gas fragmentation at high precision would be computationally prohibitive in a larger cosmological box.

The \qmarks{frozen} FDM model serves as a diagnostic tool; in this setup, we first evolve a dark matter-only FDM halo until it reaches a virialised state. We then pause the dark matter part of the simulation, thereby \qmarks{freezing} it in place. This \qmarks{frozen} state is then used to generate a static background gravitational potential. In these simulations, the gas is allowed to cool and collapse within this static cored potential. Since the dark matter potential is fixed, the gas does not experience the order-unity density fluctuations or the physical migration of the soliton centre obtained using the full-physics Schrödinger–Poisson solver. We expect that this \qmarks{frozen} potential will lead to more efficient gas accumulation than the \qmarks{dynamic} FDM case. However, we still expect to see suppression in gas infall and star formation relative to CDM due to the presence of the solitonic core.

This three-way comparison between CDM, \qmarks{dynamic} FDM, and \qmarks{frozen} FDM allows us to distinguish between effects caused by the spatial shape of the potential well (solitonic cored profile in FDM vs.\ cusped in CDM) and those caused by the dynamic wave-like oscillations and migrations of the solitonic core and surrounding interference granules. 

The range of halo masses $M_{\mathrm{h}} \approx \SIrange{e8}{e10}{\Msun}$ is chosen because it represents typical first star-forming haloes in FDM cosmologies; with this range of $m_{\mathrm{a}}$, the suppression of small-scale power means that star formation is delayed until haloes reach these mass scales. These choices are motivated by both physical relevance and numerical feasibility. Previous large-scale cosmological simulations of FDM \citep[e.\,g.][]{Schive2014, May2023, Dome2023} indicate that this halo mass range represents the bulk of first star formation in FDM cosmologies with $m_{\mathrm{a}} \sim \SI{e-22}{\eV}$. For $M_{\mathrm{h}} < \SI{e8}{\Msun}$, the FDM power spectrum cut-off typically prevents much smaller structures from forming for the given range of $m_{\mathrm{a}}$. We also aim to examine $M_{\mathrm{h}}$ values where the resulting solitonic core constitutes a significant fraction of the halo mass and so we use the approximate core–halo mass relation of \citet{Schive2014b} described in \cref{sec:theory:dynamics} to select the range of halo masses. We expect that the soliton mass fraction would correlate with the extent of any disruption to gas collapse as it dominates the central potential. This ensures that the wave-driven dynamics are a dominant gravitational force in the region where gas fragmentation and collapse occurs.

As well as computational constraints of resolving the smaller wave patterns with higher axion masses (e.\,g. $m_{\mathrm{a}} \sim \SI{e-20}{\eV}$), there is a diminishing cosmological motivation for exploring much higher $m_{\mathrm{a}}$ in the context of resolving small-scale discrepancies. At $m_{\mathrm{a}} \gtrsim \SI{e-20}{\eV}$, the FDM power spectrum and halo internal structures begin to converge toward the standard CDM limit \citep{Hu2000, Schive2014, Hui2017, Dalal2022, Powell2023}, effectively removing the features that make \qmarks{plain} FDM a distinct and appealing alternative to CDM at these mass scales. Consequently, focusing on $m_{\mathrm{a}} \sim \SI{e-22}{\eV}$ represents the most relevant region of the parameter space to explore. Our preliminary tests also showed that at \SI{e-21}{\eV}, the soliton's fluctuations are sufficiently small and high-frequency that their impact on the gas is negligible when compared to a reference CDM case within our halo mass range, which further motivates this choice of $m_{\mathrm{a}}$ range as where the most significant departures from CDM are found.

By focusing on isolated systems, we can precisely control the halo mass $M_{\mathrm{h}}$ and axion mass $m_{\mathrm{a}}$ parameters, and binding energy of our haloes. We attempted to simulate the $m_{\mathrm{a}}$ and $M_{\mathrm{h}}$ combinations in \cref{tab:simulation_parameters} that show \qmarks{---}, but due to the small periodic box and large core radii, these haloes unstable and so did not form a bound virialised halo during halo generation. This can be seen as a result of the core–halo mass relation \citep{Schive2014b, Chan2022} described in \cref{sec:theory:dynamics}, $M_{\mathrm{s}}/M_{\mathrm{h}} \propto M_{\mathrm{h}}^{-2/3}m_{\mathrm{a}}^{-1}$, which implies that for a given $m_{\mathrm{a}}$ and sufficiently low $M_{\mathrm{h}}$, the solitonic core becomes so large and diffuse that $M_{\mathrm{s}} \gtrsim  M_{\mathrm{h}}$ and thus forming a bound halo is impossible. In our simulations this occurred for our $ M_{\mathrm{h}} = \SI{8e8}{\Msun}$ halo and $m_{\mathrm{a}} \le \SI{1e-22}{\eV}$ and for our $ M_{\mathrm{h}} = \SI{3e8}{\Msun}$ halo for $m_{\mathrm{a}} \le \SI{2e-22}{\eV}$.

Further details of our simulations, including our computational setup and our initial conditions, can be found in \cref{sec:methods:code,sec:methods:ics} respectively.

\subsection{Computational Setup}
\label{sec:methods:code}

For all our simulations, we use the well-known and tested hydrodynamics code \textsc{arepo} \citep{Springel2010}, which is well-suited for resolving high-dynamic-range structures like primordial gas clouds. To model the different dark matter scenarios, we employ two distinct dark matter gravity solvers within the code.

\subsubsection{Fuzzy Dark Matter}
\label{sec:methods:code:FDM}

For the Fuzzy Dark Matter simulation, we use the \textsc{axirepo} module to simulate the FDM physics \citep{May2021, May2023}. \textsc{axirepo} makes use of a second-order pseudo-spectral \qmarks{kick-drift-kick} method to solve the SP equations \citep{Woo2009}. For this method,  the wavefunction $\psi$ and potential $V$ are discretized onto a uniform $N^3$ Cartesian grid, which allows us to use the Fast Fourier Transform (fft) and its inverse (ifft). The potential can be calculated directly from the wavefunction $\psi$ as $\rho_{\mathrm{dm}} = |\psi|^2$ and $\rho = \rho_{\mathrm{dm}} + \rho_{\mathrm{gas}}$ as follows: 
\eqn{ V = \mathrm{ifft}\sbrac{-\mathrm{fft} \sbrac{4\pi G \brac{\rho - \bar{\rho}}}/k^2},
\label{eqn:poisson-solver}}
where $k$ is the wave number for the given grid locations. The wavefunction is then evolved through a \qmarks{kick} of half a time step due to the potential:
\eqn{\psi \leftarrow \exp \sbrac{ -i \bfrac{\Delta t}{2} \bfrac{m}{\hbar} V } \psi \label{eq:kick}. }
This is followed by a full \qmarks{drift} step:
{\allowdisplaybreaks%
\begin{align}
    \hat{\psi} &\leftarrow \fft{\psi},
    \\
    \hat{\psi} &\leftarrow \exp \sbrac{-i\Delta t \bfrac{\hbar}{m} \bfrac{k^2}{2}} \hat{\psi},
    \\
    \psi &\leftarrow \ifft{\hat{\psi}}.
\end{align}%
}
The algorithm is then completed by performing another \qmarks{kick} step (\cref{eq:kick}) in order to finish the time evolution of one time step $\Delta t$.

We also ran a set of simulations where we took the same FDM initial conditions, but \qmarks{froze} the FDM. i.\,e.\ paused the SP solver described above and stopped evolving the wavefunction $\psi$ but preserved the gravitational contribution from the DM when the gas was added and evolved. This approach allows us to directly see whether any effects on the embedded gas are a result of the lack of a gravitational cusp in FDM haloes, and how much is caused by the continuous, time-varying gravitational fluctuations of the FDM.

\subsubsection{Cold Dark Matter}
\label{sec:methods:code:CDM}

Our CDM simulations treat dark matter as a collection of $256^3$ collisionless \qmarks{Monte Carlo} particles sampling phase space. Gravitational forces are calculated using the standard TreePM algorithm in \textsc{arepo}. In this approach, long-range forces are computed on a Fourier grid according to \cref{eqn:poisson-solver}, while short-range forces are handled via a hierarchical octree structure. This dual approach ensures that we capture the small-scale structure of CDM haloes (which lack the quantum pressure of FDM) with high precision \citep{Bagla2002, Springel2005}. 

To avoid numerical divergences during close-range particle encounters and to accurately capture the characteristic central \qmarks{cusp} of the CDM potential, we employ a gravitational softening length of $\epsilon = \SI{3.1}{\pc}$. We intentionally set this value to be smaller than the accretion radius of our sink particles (see \cref{sec:methods:code:sinks}). This ensures that the gravitational potential is resolved at scales below the star-formation threshold, providing a physically robust potential down to the length and density scales of our sink formation without being artificially smoothed by the softening at the scales of interest.

\subsubsection{Gas Dynamics and Gravitational Coupling}
\label{sec:methods:code:Gas}

While the dark matter is evolved on a Cartesian grid (FDM) or as \qmarks{Monte Carlo} particles (CDM), the gas component is modelled using the quasi-Lagrangian moving mesh approach of \textsc{arepo} \citep{Springel2010, Weinberger2020}. 

The gas is discretized using a Voronoi tessellation of the simulation volume, where the mesh-generating points move with the local fluid velocity. This allows the simulation to automatically increase resolution in high-density regions, such as the collapsing gas clouds at the centre of the dark matter haloes, without the numerical diffusion typical of fixed-grid Eulerian methods. 

The gravitational coupling between the dark matter and gas is handled based on the respective DM solver:
\begin{enumerate}
    \item FDM (Frozen and Dynamic): The gravitational potential $V$ is calculated on the $N^3$ Cartesian grid using the pseudo-spectral solver as described in \cref{sec:methods:code:FDM}. At each synchronization step, this potential is interpolated from the grid onto the centre of mass of each Voronoi gas cell using a standard trilinear interpolation scheme, in the same way as in a PM solver (such as the PM part of the TreePM method). The gradient of this interpolated potential then provides the gravitational acceleration for the gas due to the underlying dark matter. The rest of the gravitational contribution from the gas is done as in the CDM case (using the TreePM algorithm, see below). 
    \item CDM: The gravitational acceleration for each gas cell is calculated directly by summing the contributions from the dark matter particles using the TreePM algorithm \citep{Springel2010}. The tree provides the short-range forces from nearby particles, while the Fourier grid handles the long-range component, ensuring that the gas \qmarks{sees} the same high-resolution potential, including the central cusp, as the CDM particles.
\end{enumerate}

Alongside the gravitational considerations, we also employ the sink particle prescription as detailed by \citet{Bate1995, Federrath2010, Tress2020, Wollenberg2020} (further details in \cref{sec:methods:code:sinks}) and the primordial chemistry network of \citet{Clark2011} (details in \cref{sec:methods:code:chemical_model}) . We set the softening scale of the gas at $\SI{1.9e-1}{\pc} = r_\mathrm{sink}/16$ which is the minimum size a gas cell should be able to reach based off the sink creating radius (\cref{sec:methods:code:sinks}) and Jeans refinement criteria requiring 16 cells per jeans length.

\begin{figure*}
    \centering
    \includegraphics[width=\linewidth]{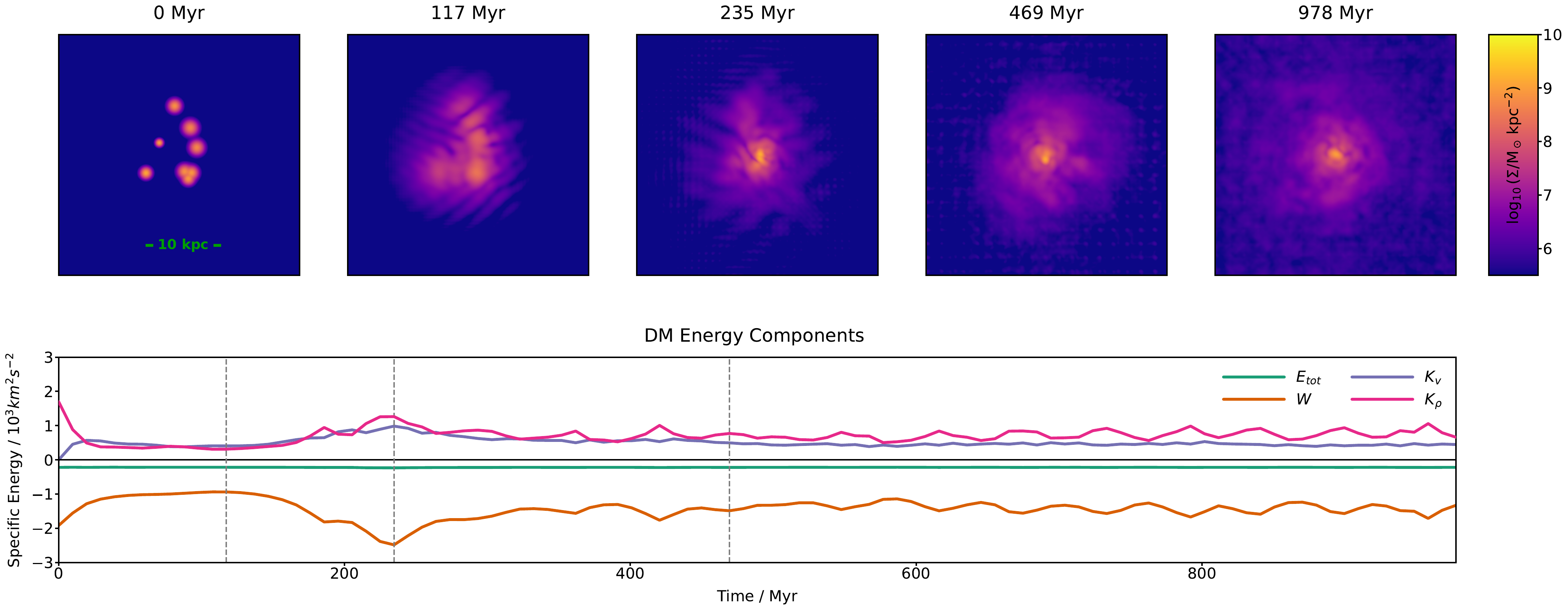}
    \caption{Top: Projected (log) density snapshots of the halo assembly (DM only) starting from the initial conditions (the leftmost panel), followed by three intermediate steps at \SIlist{117; 235; 469}{\Myr}, taken from our $\SI{3e9}{\Msun}$ halo and with $m_\mathrm{a} = \SI{1e-22}{\eV}$ The final halo used for our simulations is shown in the rightmost panel. Bottom: The total energy of the system (green, $E_{\mathrm{tot}}$) and energy components: potential energy (orange $W$), classical kinetic energy (purple $K_v$) and quantum kinetic energy (pink $K_\rho$) throughout the assembly timeline of the halo. The vertical dashed lines correspond to the three instances shown in the intermediate top panels. The total energy is clearly conserved throughout the simulation box. We highlight the horizontal axis at zero to show clearly that the total energy of our haloes is less than zero, i.\,e. the haloes are bound.}
    \label{fig:fdm_assembly}
\end{figure*}

\subsubsection{Sink Prescription}
\label{sec:methods:code:sinks}

In order to limit the indefinite refinement of the gas cells and to introduce an easy measure of the quantity and distribution of cold dense gas we use a sink particle prescription within \textsc{arepo}. In our simulations, these sink particles serve as a proxy for star-forming regions within the halo where the gas has met the requirements to continue cooling and collapse as outlined below \citep{Schauer2021}. The sink particle implementation we use here is the same as that introduced in \citet{Tress2020, Tress2021} and \citet{Wollenberg2020} based off \citet{Federrath2010}. These sink particles are formed when a gas cell that is at a local minimum of the potential satisfies the following criteria:
\begin{enumerate}
    \item the gas cell has a number density $n > n_\mathrm{sink} = \SI{e2}{\per\cm\cubed}$;
    \item the local velocity field is convergent;
    \item the local acceleration is convergent;
    \item the total energy of the gas within the sink accretion radius ($r_{\mathrm{sink}}$) is negative, i.\,e.\ the gas is gravitationally bound;
    \item there are no other sink particles within $r_{\mathrm{sink}}$. 
\end{enumerate}
These sink particles act as point masses within the simulation which are able to accrete mass from any gas cells around them that are within the sink particle's accretion radius $r_{\mathrm{sink}}$.
We set the value of $r_{\mathrm{sink}}$ to be comparable to the Jeans length at the sink formation density threshold $n_{\mathrm{sink}}$. In our simulations, gas at this density typically has a temperature of $\sim \SI{e2}{\K}$, resulting in a Jeans length of around \SI{3}{\pc}, and so we set $r_{\mathrm{sink}} = \SI{3}{\pc}$.
Gas within a distance $r_{\mathrm{sink}}$ of an existing sink is accreted if it meets the following requirements: it must be gravitationally bound to the sink and must have a density above $n_{\mathrm{sink}}$. 

Our choice of value for $n_{\mathrm{sink}}$ is motivated by our desire to effectively capture the bulk properties of the active, star-forming region within the halo centre while at the same time ensuring that our simulations remain computationally efficient. 

It is important to clarify that at this density, the sinks do not represent individual Population III stars. Instead, they serve as numerical proxies for the onset of the runaway cooling phase where the free fall ($t_\mathrm{ff})$ and cooling times ($t_{\mathrm{cool}}$) of the gas are much less than characteristic timescale of any of the DM dynamics $\tau_{\mathrm{osc}}$. These particles accurately capture the relative integrated star formation rate of the halo, and can be used to compare the accumulation of dense, star-forming gas between haloes, even if they don't resolve individual stars \citep{Gnedin2009, Jeon2012, Schauer2021}. By using these sink particles, we intentionally circumvent the need to model the full fragmentation and collapse into individual stellar objects, a process that would otherwise necessitate resolving significantly smaller scales and much higher densities, and hence would have much greater computational requirements.

\subsubsection{Chemical Model}
\label{sec:methods:code:chemical_model}

In metal-free primordial gas, radiative cooling is governed by a small number of atomic and molecular processes, with the dominant cooling channel depending primarily on halo mass and virial temperature. In \qmarks{minihaloes} ($M_{\mathrm{h}} \sim \SIrange{e5}{e7}{\Msun}$, $T_{\mathrm{vir}} \lesssim \SI{e4}{\K}$), cooling is enabled by molecular hydrogen (H$_2$), which forms through gas-phase reactions and cools the gas via rotational-vibrational transitions, allowing collapse to temperatures of $\sim \SI{200}{\K}$ \citep{Abel2002, Bromm2004, Glover2008, Bromm2011}. In more massive haloes like the ones we aim to study ($M_{\mathrm{h}} \gtrsim \SIrange{e7}{e8}{\Msun}$ at $z \sim \numrange{10}{20}$), the virial temperature exceeds $\sim \SI{e4}{\K}$ and cooling is instead dominated initially by collisional excitation of atomic hydrogen and subsequent Ly$\alpha$ emission \citep{Oh2002, Wise2007, Bromm2011}. In these so-called atomic-cooling haloes, cooling times are typically much shorter than the dynamical time, making gas inflow and central condensation highly efficient. While atomic hydrogen cooling regulates the onset of collapse in such systems, molecular hydrogen becomes important at higher densities, where it influences the thermodynamic evolution and fragmentation properties of the gas \citep{Glover2008, Greif2012}.

The primary motivation for our choice of chemical model is to accurately track the transition from virialised gas to the first star-forming regions. We adopted a primordial chemical network and cooling function based on the ones introduced in \citet{Clark2011}, with updates as described in \citet{Schauer2017}. The model handles the primordial gas chemistry of 12 species ($\mathrm{H}$, $\mathrm{H}^+$, $\mathrm{H}^-$, $\mathrm{H}_2^+$, $\mathrm{H}_2^{}$, $\mathrm{He}$, $\mathrm{He}^+$, $\mathrm{He}^{++}$, $\mathrm{D}$, $\mathrm{D}^+$, $\mathrm{HD}$, and free electrons) via 45 chemical reactions.

While \textsc{arepo} handles the standard hydrodynamics, this chemistry module is essential for Population~III studies as it solves the non-equilibrium rate equations for each species alongside the hydrodynamical evolution. The primary cooling channels in our simulations are driven by $\mathrm{H}_2$ and $\mathrm{HD}$ molecules. These represent the dominant cooling mechanisms for primordial gas at temperatures below \SIrange{5000}{8000}{\K}, where atomic cooling becomes inefficient \citep{Glover2008}. In particular, $\mathrm{HD}$ cooling is critical for allowing the gas to surpass the $\mathrm{H}_2$ cooling limit and reach temperatures below \SI{200}{\K} \citep{Galli1998}, a regime that is vital for determining the fragmentation scale of the gas. The network further accounts for ionization and recombination, three-body $\mathrm{H}_2$ formation, and the thermal effects of chemical composition changes, shocks, and adiabatic compression or expansion.

At each time step, the chemical abundances are updated based on the local density and temperature of the Voronoi cell. The cooling rate and heating rate of the gas cells are updated simultaneously. Crucially, the local chemical state determines the mean molecular weight $\mu$, which is used to calculate the gas temperature $T$ from the internal energy $u$ via the equation of state:
\eqn{ T = \frac{(\gamma - 1) u \mu m_{\mathrm{p}}}{k_{\mathrm{B}}}, }
where $m_{\mathrm{p}}$ is the proton mass, $\gamma$ is the adiabatic index, $u$ is the specific internal energy, and $k_{\mathrm{B}}$ is the Boltzmann constant. This feedback loop ensures that the collapse of the gas is correctly regulated. For the simulations presented here, we set $\gamma = 5/3$ since the gas is dominated by monatomic particles ($\mathrm{H}$ and $\mathrm{He}$ in neutral regions or electrons and ions in ionized regions) at all of the densities we resolve.

\subsection{Initial Conditions}
\label{sec:methods:ics}

Below we detail how our initial conditions were generated and the motivations for our methods.

\subsubsection{Halo Generation}
\label{sec:methods:ics:halo_generation}

In order to isolate the impact of FDM small-scale dynamics from the differences in large-scale structure formation, we initialize our haloes using a non-cosmological, dark matter-only, solitonic core merger setup, similar to that of \citet{Mocz2017}. This allows us to generate CDM and FDM haloes with identical global properties, specifically total halo mass, total energy, and angular momentum. In full cosmological volumes, the FDM suppression of small-scale power naturally leads to different halo histories, and therefore makes comparing the gas processes and collapse between haloes of different dark matter models difficult to disentangle from the effects of different formation histories. In contrast, with our controlled setup we can isolate any effect that the bulk halo properties might have when comparing our FDM, \qmarks{frozen} FDM, and CDM cases.

For each halo generation (both FDM and CDM), we place a set of solitonic cores within our box and allow them to merge, an example of which can be seen in \cref{fig:fdm_assembly}. It shows a series of snapshots starting with our initial solitonic cores, and various stages of their merger history, with the final halo shown in the rightmost snapshot. It should be noted that for our simulation boxes, due to the periodic boundary conditions and small overall box size, the virial ratio $2K/|W| $ is greater than one despite the system appearing to reach a steady state. This is likely due to the underestimation of $|W|$ as has been noted in small periodic boxes before \citep{Power2006, Rios2023, Blum2025}.

We evolved these dark matter-only systems for between \SIrange{1}{5}{\Gyr} depending on the halo mass. The initial assembly phase was monitored and stopped when the dark matter potential has reached a stable, virialised equilibrium. This \qmarks{settling} phase is monitored via the energy evolution (\cref{fig:fdm_assembly}); we ensure the classical kinetic ($K_v$) energy has plateaued, indicating that the initial \qmarks{ringing} of the merger has subsided. In FDM haloes, persistent exchange between $K_\rho$ and $W$ is expected post-virialisation due to the large wave effects and the Heisenberg uncertainty principle acting on the wavefunction, see \citet{Mocz2017}. The lower panel of \cref{fig:fdm_assembly} shows how the energy components, calculated as in \cref{eqn:energy}, change during this merger process, the initial \qmarks{ringing} of the merger at $\sim \SI{235}{\Myr}$ and subsequent stabilization of $K_v$. It was only once the initial halo had been assembled that we then stopped the SP solver in order to create our \qmarks{frozen} FDM haloes. 

It should be noted that we attempted to run the three combinations of low $m_{\mathrm{a}}$ and low $M_{\mathrm{h}}$ marked with \qmarks{---} in \cref{tab:simulation_parameters} but due to the low mass of these haloes, the periodic boundary conditions, and their low $m_{\mathrm{a}}$, these setups resulted in diffuse, spatially extended DM distributions that did not form distinct haloes. We also did not run $m_{\mathrm{a}} = \SI{2e-22}{\eV}$ for our highest halo mass for reasons of computational efficiency as we did not expect to see any interesting insights that would not be clear from the $m_{\mathrm{a}} = \SI{1e-22}{\eV}$ or \SI{3e-22}{\eV} cases.

\subsubsection{Initial State of the Gas}
\label{sec:methods:ics:gas}

Once the dark matter halo is virialised, gas is added as a uniform background, with an initial total of $256^3$ gas cells and a total mass consistent with the cosmic baryon fraction. The gas is initially at rest. While a large-scale relative velocity between DM and gas (streaming velocity) of (on average) $\sim \SI{30}{\km\per\s}$ is expected at recombination which decays by a factor of $(1 + z)$ leading to a relative velocity of $v_{\mathrm{bc}} \approx \SI{0.6}{\km\per\s}$ at $z = 20$ \citep{Tseliakhovich2010}, we omit this effect in this work, as well as any intrinsic turbulence/kinetic energy in the initial gas state, to purely isolate the local impact of FDM dynamics. An initial comparison of the velocity of the soliton's random walk at the centre of the halo, which is a few tens of \si{\km\per\s}, further indicates that any physically meaningful $v_{\mathrm{bc}}$ would be negligible in comparison to the effects of the soliton's dynamics. We hope to address these considerations in future work. The initial gas temperature is set to \SI{e4}{\K}, representing the typical virial temperature of atomic cooling haloes at $z \sim 20$ \citep{Schauer2021, Kulkarni2021, Dome2024}. The chemical network is initialized with primordial abundances (\cref{tab:chem}), with a small seed fraction of $\mathrm{H}_2$ (\num{2e-6}).

\begin{table}

\centering
    \caption{Initial chemical species abundances, defined relative to the number of H nuclei. Since we only set the overall deuterium (D) abundance first, then the D$^+$ and HD abundances, the abundance of all deuterium species together are shown together in the top row.}
    \begin{tabular}{lS[table-format=1.1e+1]}
    \toprule
    Species & {Initial Abundance $n_\mathrm{species}/n_{\mathrm{H}}$} \\
    \midrule
    D + D$^{+}$ + HD & 2.6e-5 \\
    \midrule
    H$^+$            & 1.0e-4 \\
    $\mathrm{H}_2$   & 2.0e-6 \\
    H$^-$            & 0.0 \\
    H$_2^+$          & 0.0 \\
    D$^+$            & 2.6e-9 \\
    HD               & 0.0 \\
    He               & 7.9e-2 \\
    He$^+$           & 0.0 \\
    He$^{++}$        & 0.0 \\
    Free electrons ($e^-$) & 1.0e-4 \\
    \bottomrule
    \end{tabular}
    \label{tab:chem}
\end{table}

\section{Results}
\label{sec:results}

We begin by examining the fundamental properties of the FDM haloes we have generated, which serve as the foundation for all subsequent gravitational effects on the baryons. Unlike the standard CDM paradigm, where haloes are described by a smooth, cusped potential, FDM haloes are characterized by a central solitonic core and persistent, ubiquitous density fluctuations as described in \cref{sec:theory:dynamics}. These fluctuations are a direct consequence of the wave-like nature of the FDM particles, leading to a gravitational potential that is not static but rather constantly in motion. Here we explore the impact of these fluctuations on the gravitational potential, the collapse of the primordial gas and the subsequent star formation.

\subsection{Gravitational Potential and Density Fluctuations at the Halo Centre}
\label{sec:potential}

It is well established that the shape of the potential wells of CDM and FDM haloes differ, especially below the core radius $r_{\mathrm{c}}$. As described in \cref{sec:theory:FDM}, while CDM haloes are characterized by \qmarks{cuspy} destiny profiles at small radii, FDM haloes form a solitonic core at their centre \citep{Hu2000, Mocz2017, May2021, Nori2024}.

\begin{figure}
    \centering
    \includegraphics[width=\linewidth]{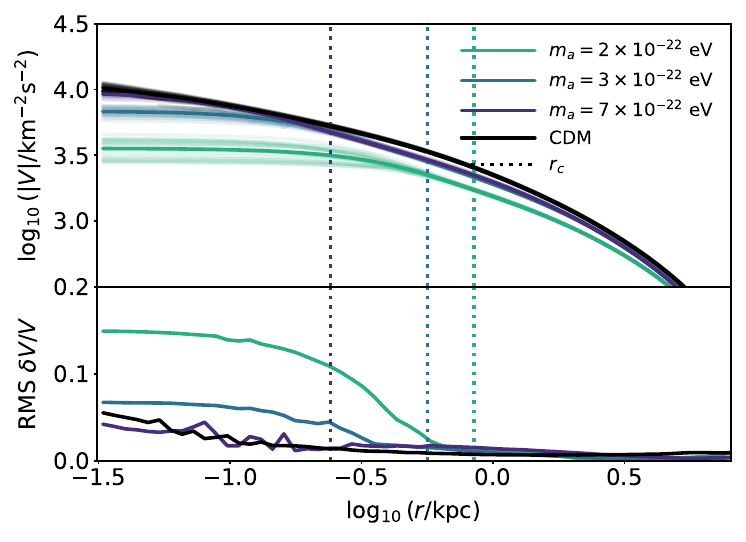}
    \caption{Top: The time-averaged gravitational potential profiles of FDM haloes for different axion masses ($m_{\mathrm{a}} = \SIlist{2e-22; 3e-22; 7e-22}{\eV}$) and the corresponding CDM halo, all with a mass of $M_{\mathrm{h}} = \SI{8e8}{\Msun}$. Individual snapshots are shown by the translucent lines. The vertical dotted lines indicate the core radius ($r_{\mathrm{c}}$) for each FDM halo. The potential flattens inside the core radius for FDM haloes, a signature of quantum pressure. Bottom: The root-mean-square (RMS) fluctuation of the potential ($\delta V/ V$) for the same haloes. This quantifies the fluctuations of the solitonic core at the centre of each FDM halo. The fluctuations are most pronounced and extend to larger radii for smaller $m_{\mathrm{a}}$, and they decrease significantly outside the solitonic core radius.}
    \label{fig:potential_profiles}
\end{figure}

\Cref{fig:potential_profiles} shows the spherically averaged gravitational potential profile and the root-mean-square (RMS) fluctuations of the potential for the $M_{\mathrm{h}} = \SI{8e8}{\Msun}$ haloes, comparing FDM cases with varying $m_{\mathrm{a}}$ to the CDM case. We choose this halo mass as the contract in behaviour between the CDM and lightest $m_\mathrm{a}$ FDM cases was most pronounced (See \cref{sec:results:gas_accumulation} for further details on the halo mass dependence of our results). The upper panel displays the time-averaged potential profiles in bold, with the potential profiles for each snapshot shown as translucent lines. For each FDM halo, a core is evident at small radii, where the potential flattens due to the quantum pressure. As the axion mass $m_{\mathrm{a}}$ increases, the FDM potential profile transitions to become more like the CDM profile, which has a sharper cusp. The potential of the $m_{\mathrm{a}} = \SI{7e-22}{\eV}$ halo is very similar to the CDM case, suggesting a reduced impact on gravitational dynamics for higher axion masses.

The lower panel of \cref{fig:potential_profiles} shows the dimensionless RMS fluctuation of the potential ($\delta V/ V$) as a function of radius. These fluctuations occur on a timescale of $\sim \tau_{\mathrm{osc}}$ as outlined in \cref{sec:theory:dynamics} and are a key signature of FDM haloes, resulting from the persistent interference of the dark matter wavefunction. The plot shows a clear dependence of these fluctuations on the axion mass. For the lightest axion mass $m_{\mathrm{a}} = \SI{2e-22}{\eV}$ that we ran for our $\SI{8e8}{\Msun}$ halo, the RMS fluctuation is notably high at the centre of the halo, reaching a value of $\sim 0.15$ and extending out to a radius corresponding to the soliton core radius ($r_{\mathrm{c}}$). As $m_{\mathrm{a}}$ increases, the magnitude of these fluctuations decreases significantly, with the $m_{\mathrm{a}} = \SI{7e-22}{\eV}$ case showing very low variation, comparable to the CDM case. It is worth noting that there are persistent granular fluctuations throughout the halo, but these have a minimal impact on the spherically averaged gravitational potential at radii larger than the soliton radius because they are at much lower density than the soliton itself, and as a result the soliton's properties dominate the shape of the potential well.

The persistent and significant fluctuations in the central gravitational potential, particularly for low $m_{\mathrm{a}}$, are the primary mechanism through which FDM dynamics affect the baryonic gas. These fluctuations act as a source of energy and turbulence for the gas, hindering its efficient collapse and leading to the effects we discuss in the following subsections.

\subsection{Distribution of Cold Dense Gas and FDM Dynamics as a Driver of Turbulence }
\label{sec:results:gas_distribution}

\begin{figure*}
    \centering
    \includegraphics[width=0.9\linewidth]{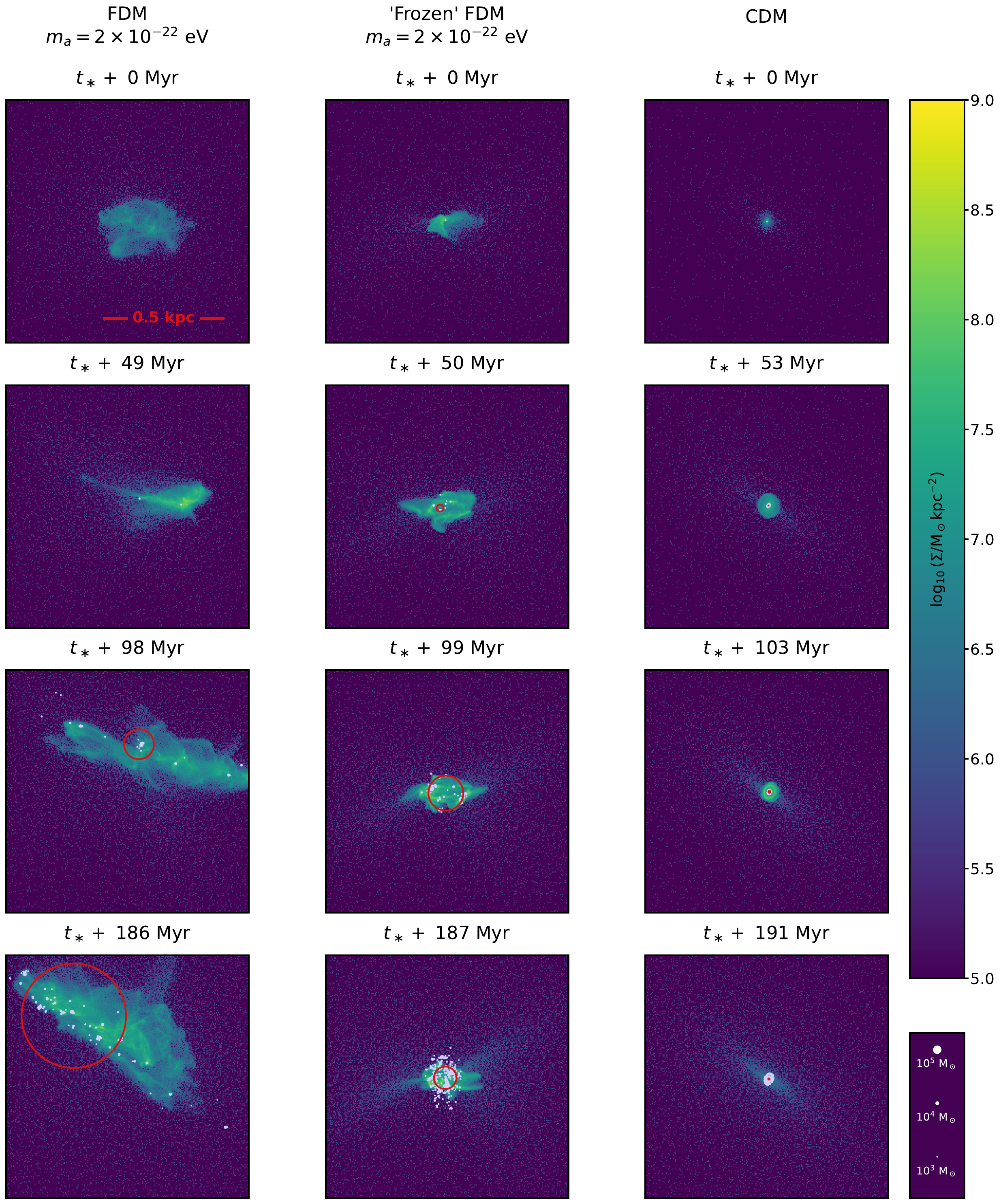}
    
    \caption{Gas density projections in the central region of each halo for a selection of our simulations with a halo mass of \SI{8e8}{\Msun}: FDM with $m_{\mathrm{a}} = \SI{2e-22}{\eV}$ (left), the \qmarks{frozen} FDM with the same $m_{\mathrm{a}} = \SI{2e-22}{\eV}$ (centre), and a CDM case (right). The first row shows the gas state at the formation of the first sink. As the simulation proceeds, in the FDM cases, the gas is distributed more widely across the central region of the halo as it is affected by the dynamics of the central soliton. In the CDM case, gas remains tightly bound in the central region. The gray dots show the location of the sink particles, with the radius scaled by the sink mass $m_\mathrm{sink}^{1/3}$. The red circle shows the half-mass radius of the sink particles. There is no half mass radius shown for the \qmarks{dynamic} FDM case for $t_\ast + \SI{49}{\Myr}$ as there was still only one sink formed at this time.}
    \label{fig:snapshot_series}
\end{figure*}

We now examine how the unique central dynamics of FDM haloes physically restructure the gas distribution. \Cref{fig:snapshot_series} presents a time-series of gas density snapshots for a halo of $M_{\mathrm{h}} = \SI{8e8}{\Msun}$, comparing the \qmarks{dynamic} FDM with $m_{\mathrm{a}} = \SI{2e-22}{\eV}$ to the corresponding \qmarks{frozen} FDM case and the control CDM scenario. The snapshots are shown starting from the moment of the first sink formation $t_\ast$ followed by snapshots at $\Delta t \approx \SI{50}{\Myr}$, $\Delta t \approx \SI{100}{\Myr}$ and $\Delta t \approx \SI{190}{\Myr}$ since $t_\ast$. We recall that sink particles, shown in gray in \cref{fig:snapshot_series}, represent gas that has entered irreversible runaway collapse (see \cref{sec:methods:code:sinks}), and thus provide a proxy for star formation efficiency.

While the CDM case (right column) exhibits a rapid, centrally-concentrated collapse into a single dense region, both FDM cases show a more extended gas morphology. However, a critical distinction emerges between the two FDM models: in the \qmarks{dynamic} case (left column), the gas distribution is significantly more extended and fragmented than in the \qmarks{frozen} case (middle column). Although the \qmarks{frozen} model possesses the same flattened gravitational potential (the soliton core) as the full FDM scenario, the gas settles into a relatively symmetric, albeit wider than in the CDM case, configuration. In contrast, the \qmarks{dynamic} FDM potential exerts stochastic gravitational torques that continuously stir the gas. This dynamic stirring drives turbulence that counteracts the inward gravitational pull, maintaining a diffuse cloud that becomes increasingly spatially extended over time. 

This physical disruption is further quantified by the evolution of the sink particle half-mass radius, represented by the red circles in \cref{fig:snapshot_series}. This half-mass radius is the radius of a sphere around the centre of mass of sinks that contains half of the total mass in sinks.  In the CDM halo, sinks form a compact central cluster with a stable, small half-mass radius, indicating efficient and localized gas collapse. Conversely, in \qmarks{dynamic} FDM, the sinks are found far from the halo centre; they are distributed across a wider volume, and their half-mass radii often increase as the simulation progresses. This dramatic increase in the half-mass radius is a direct signature of the FDM soliton's dynamics.

The fact that the half-mass radius in the \qmarks{frozen} case remains significantly smaller than in the \qmarks{dynamic} case (though larger than CDM) confirms that while the core geometry sets a floor for the spatial extent of star formation, the fluctuation dynamics are the primary driver of the observed distribution. This implies that the spatial clustering of the first stars in FDM cosmologies may be much lower than predicted by static halo models, potentially impacting the local feedback, binary formation, and subsequent galaxy assembly.

\begin{figure}
    \centering
    \includegraphics[width=\linewidth]{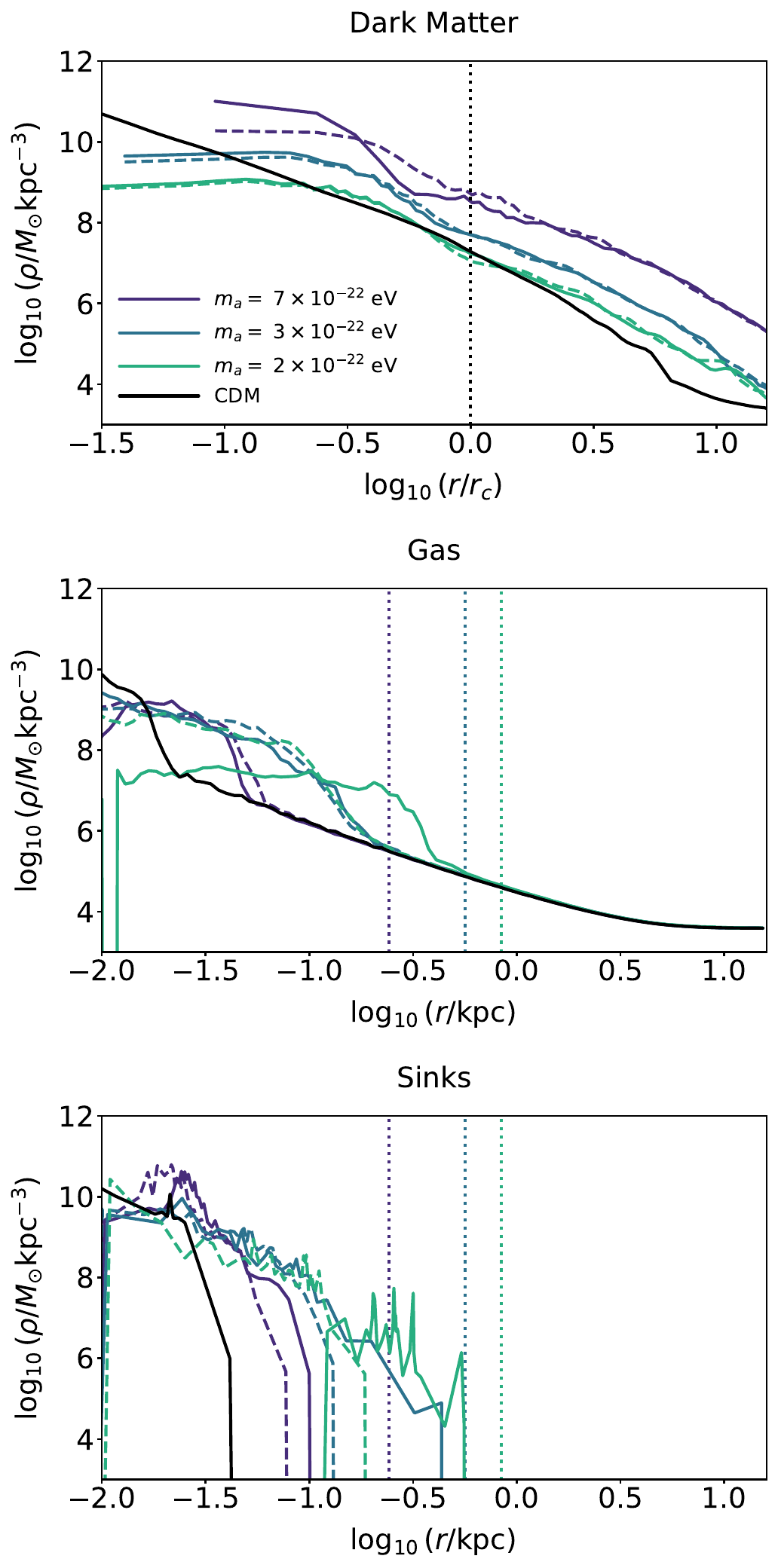}

    \caption{Radial density profiles for the \SI{8e8}{\Msun} halo at a late-time snapshot with $t \approx \SI{250}{\Myr}$ from the start of the simulations. The dashed lines show the \qmarks{frozen} FDM cases while the dotted lines show the core radius $r_{\mathrm{c}}$. Top: Dark matter density profiles normalized by the core radius $r_{\mathrm{c}}$. For completeness we show CDM normalized with $r_{\mathrm{c}} = \SI{1}{\kpc}$. The FDM cases exhibit a flattened solitonic core, while the CDM shows a characteristic cusp. Middle: Gas density profiles.  Bottom: Radial distribution of sink particles. In the FDM simulations, particularly at lower axion masses, the sinks are spatially dispersed and displaced from the halo centre. The vertical dotted lines indicate the core radius $r_{\mathrm{c}}$ for each respective FDM model.}
    \label{fig:density_profiles}
\end{figure}

The impact of these dynamics is further quantified in \cref{fig:density_profiles}, which displays the spherically averaged radial density profiles for dark matter, gas, and sinks in the \SI{8e8}{\Msun} halo. In agreement with literature, our dark matter profiles (top panel), which have been normalized by the core radius $r_{\mathrm{c}}$, clearly distinguish the FDM models from the CDM cuspy central density by the presence of a central solitonic core.

The gas density profiles (middle panel) reveal a complex relationship between the dark matter potential and the efficiency of gas collapse. In the CDM case, the central gas density is lower than in the FDM models up to very small radii $r \lesssim 10^{-1.5} \: r_{\mathrm{c}}$. It is only below this small radius where the gas density increases above the level seen in the FDM haloes. This is due to the highly efficient accretion of gas onto the few sink particles that form at the centre of the halo, which depletes the central gas reservoir. In the FDM cases, however, the gas more closely tracks the shape of the solitonic core and remains at high densities out to larger radii as $m_{\mathrm{a}}$ is decreased. This suggests that while the gas is able to accumulate within the core, in the FDM case the gas is prevented from completing the transition into runaway collapse into sinks as efficiently as in CDM. This is further discussed in \cref{sec:results:gas_accumulation}. 

A comparison between the FDM models shows that the central gas density is suppressed for the lowest axion masses, with the $m_{\mathrm{a}} = \SI{2e-22}{\eV}$ case showing a lower central peak than either the $m_{\mathrm{a}} = \SI{7e-22}{\eV}$, $m_{\mathrm{a}} = \SI{3e-22}{\eV}$ or any of the \qmarks{frozen} models. This confirms that the injection of energy and turbulence by the halo dynamics is more pronounced at lower axion masses, and actively inhibits the gas from reaching the densities required for sink formation for the lowest axion masses, even when a substantial gas reservoir is present.

Finally, the radial distribution of sink particles (bottom panel) illustrates the spatial displacement of the sinks from the halo centre. The sinks in the \qmarks{dynamic} FDM models are scattered across a wider volume when compared to their \qmarks{frozen} counterparts, particularly for $m_{\mathrm{a}} = \SI{2e-22}{\eV}$. This larger-scale distribution is a hallmark of the FDM dynamics; as the gravitational potential minimum shifts stochastically, the sites of gas collapse are never fixed, leading to a fragmented and non-centralized distribution of the sinks within our simulations.

\subsection{Angular Momentum and Rotational Support}
\label{sec:results:angular_momentum}

We might expect that the dynamics at the centre of the FDM haloes can induce torques on the gas if the collapse timescale is longer than that of the dynamical timescale of the FDM fluctuations, and hence lead to a degree of enhanced rotational support against collapse. This can be quantified by examining the specific angular momentum ($j$) of the baryonic matter (gas plus sinks) as a function of radius in each of our haloes. We can compare this to the specific angular momentum required to support the gas against collapse at a given radius. For any given mass element, its specific angular momentum around a given axis is given by $\mathbf{j} = \mathbf{r} \times \mathbf{v}$. We can then link this to a centripetal force at a  given radius $r$ via
\eqn{F_{\mathrm{cent}} = \frac{v_\phi^2}{r} = \frac{j(r)^2}{r^3}, }
where $v_\phi$ is the velocity perpendicular to the radial vector, and $j(r) = |\mathbf{j}(r)| = rv_\phi$ is the magnitude of the total specific angular momentum of a spherical shell at radius $r$. We can then link this force to the gravitational force at a given radius to find an estimate for the angular momentum required to support the gas as a function of radius via $F_{\mathrm{cent}} = -F_{\mathrm{grav}} = \nabla V$, thus arriving at
\eqn{\frac{j_{\mathrm{crit}}(r)^2}{r^3} \approx \pdiff{V}{r}\Rightarrow j_{\mathrm{crit}}(r) \approx \sqrt{\pdiff{V}{r} r^3 }.}

\begin{figure}
        \centering
        \includegraphics[width=\linewidth]{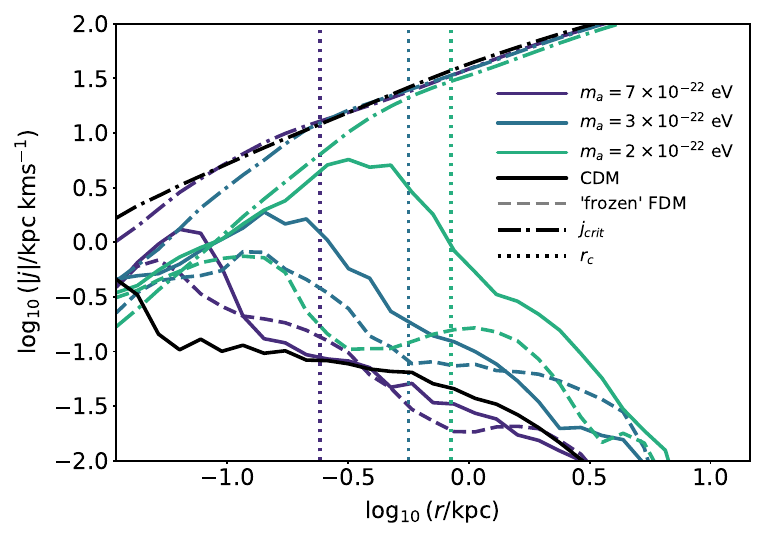}
        \caption{Time-averaged specific angular momentum ($j$) of gas and sinks as a function of radius ($r$) from the halo centre. The solid black line represents the CDM case, solid coloured lines represent FDM cases with different axion masses ($m_{\mathrm{a}}$, see legend), while dashed lines correspond to the \qmarks{frozen} FDM cases. The dash-dotted lines indicate the critical angular momentum ($j_{\mathrm{crit}}$) required for rotational stabilization against gravitational collapse. Vertical dotted lines denote the theory value for the soliton radius ($r_{\mathrm{c}}$) for each FDM halo mass \citep{Schive2014}. Note the significant induced rotation in FDM cases (especially for lower $m_{\mathrm{a}}$) below $r_{\mathrm{c}}$, where $j$ approaches or exceeds $j_{\mathrm{crit}}$, in contrast to the minimal induced rotation in the CDM and the \qmarks{frozen} FDM cases. All data are taken from the \SI{8e8}{\Msun} halo mass case at a late evolutionary time, with the snapshots taken from more than \SI{100}{\Myr} after the formation of the first sink.}
        \label{fig:angular_momentum}
\end{figure}

\Cref{fig:angular_momentum} shows the specific angular momentum $j$ calculated for each spherical shell for the combined gas and sinks as a function of $r$ from the halo centre for both CDM and various FDM cases, alongside the critical angular momentum, $j_{\mathrm{crit}}$, required for rotational stabilization against collapse (shown with dot-dashed lines in \cref{fig:angular_momentum}). The key feature of these data is the significant amount of rotation induced within the gas in our \qmarks{dynamic} FDM haloes, particularly for radii below the soliton radius, $r_{\mathrm{c}}$, and for lighter axion masses, $m_{\mathrm{a}}$. 

\begin{figure*}
        \centering
        \includegraphics[width=\linewidth]{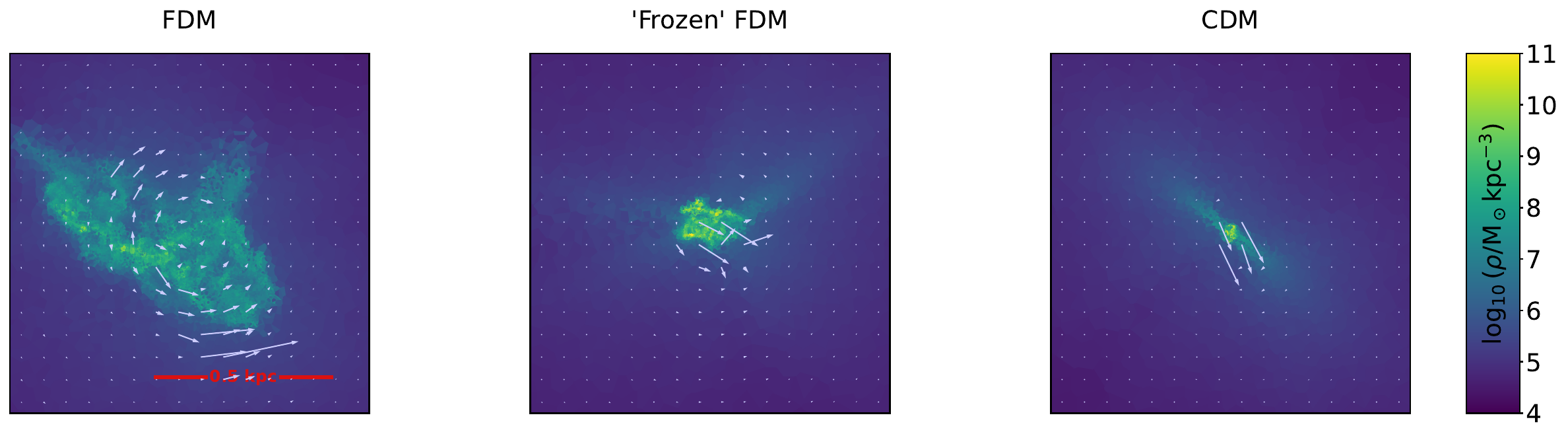}
        \caption{Gas density slices through the halo centre with projected angular momentum vectors for a late-time snapshot, synchronized with the last row of \cref{fig:snapshot_series} at $t \approx t_\ast + \SI{190}{\Myr}$. We show the \qmarks{dynamic} FDM with $m_{\mathrm{a}} = \SI{2e-22}{\eV}$ (left), the corresponding \qmarks{frozen} FDM (middle), and the CDM case (right) for a halo mass of \SI{8e8}{\Msun}. All panels have the same spatial extent and scale as the leftmost panel. The vectors are scaled by the square root of their magnitude to highlight its large range.}
        \label{fig:angular_momentum_projection}
\end{figure*}

This rotation is visualized in \cref{fig:angular_momentum_projection}, which shows the vector projections of the specific angular momentum of the gas (with arrows scaled by the square root of their magnitude for better visualization) for $M_{\mathrm{h}} = \SI{8e8}{\Msun}$ and $m_{\mathrm{a}} = \SI{2e-22}{\eV}$ at a late-time snapshot synchronized with the last row of \cref{fig:snapshot_series} at $t \approx t_\ast + \SI{190}{\Myr}$. In the \qmarks{dynamic} FDM case (left), the gas has significant angular momentum out to far larger radii. This \qmarks{stirring} is a direct consequence of the solitonic core's random walk and density fluctuations, which exert time-varying gravitational torques on the surrounding gas. In contrast, the spatial extent of the region with high angular momentum is much smaller in the CDM case, with gas mostly following a direct, radial path into the central potential minimum. 

For the \qmarks{dynamic} FDM cases, especially $m_{\mathrm{a}} = \SI{2e-22}{\eV}$ and $m_{\mathrm{a}} = \SI{3e-22}{\eV}$, the specific angular momentum $j$ of the gas within the central regions ($r \lesssim r_{\mathrm{c}}$) is notably higher compared to the CDM and \qmarks{frozen} FDM counterparts (as shown in \cref{fig:angular_momentum}). This trend suggests that the induced rotation in FDM cases is a direct consequence of the dynamic nature of the solitonic core. The persistent random walk and density fluctuations of the FDM soliton exert time-varying gravitational torques on the surrounding gas, imparting angular momentum to it. This effect is more pronounced for lighter axion masses because the de Broglie wavelength is larger, leading to more extended and hence gravitationally influential solitonic cores and a greater effect on the gas out to larger radii. Furthermore, the amplitude of these density fluctuations also scales with $\lambda_{\mathrm{dB}}$, as the larger coherence length in lighter axions results in more massive, slower-moving interference granules that exert stronger and more persistent influence on the solitonic core.

Crucially, in the \qmarks{dynamic} FDM cases, this specific angular momentum reaches or exceeds the critical threshold ($j_{\mathrm{crit}}$) required to rotationally stabilize the gas against collapse (dot-dashed lines, \cref{fig:angular_momentum}). While the \qmarks{frozen} FDM results (\cref{sec:results:gas_accumulation}) demonstrate that a static cored potential is sufficient to delay the initial timing of gas collapse, the dynamic treatment reveals a secondary, structural effect: the gas is prevented from reaching the same high-density, compact configurations seen in the static cases. Instead, the injection of angular momentum and turbulence leads to a more spatially extended gas distribution. In contrast, the \qmarks{frozen} FDM and CDM cases show negligible induced rotation, allowing gas to settle more efficiently into the potential minimum. This highlights a key physical distinction: whereas the soliton core geometry sets the timing of star formation, the soliton dynamics dictate the spatial scale of the resulting dense gas structures. By identifying this rotational stabilization (a mechanism currently absent from analytical models of FDM-driven star formation), we suggest that the suppression of gas efficiency in FDM haloes is likely greater than previously estimated.

\subsection{Accumulation of Gas in Sinks: Core vs. Dynamics}
\label{sec:results:gas_accumulation}

\begin{figure*}
    \centering
    \includegraphics[width=\linewidth]{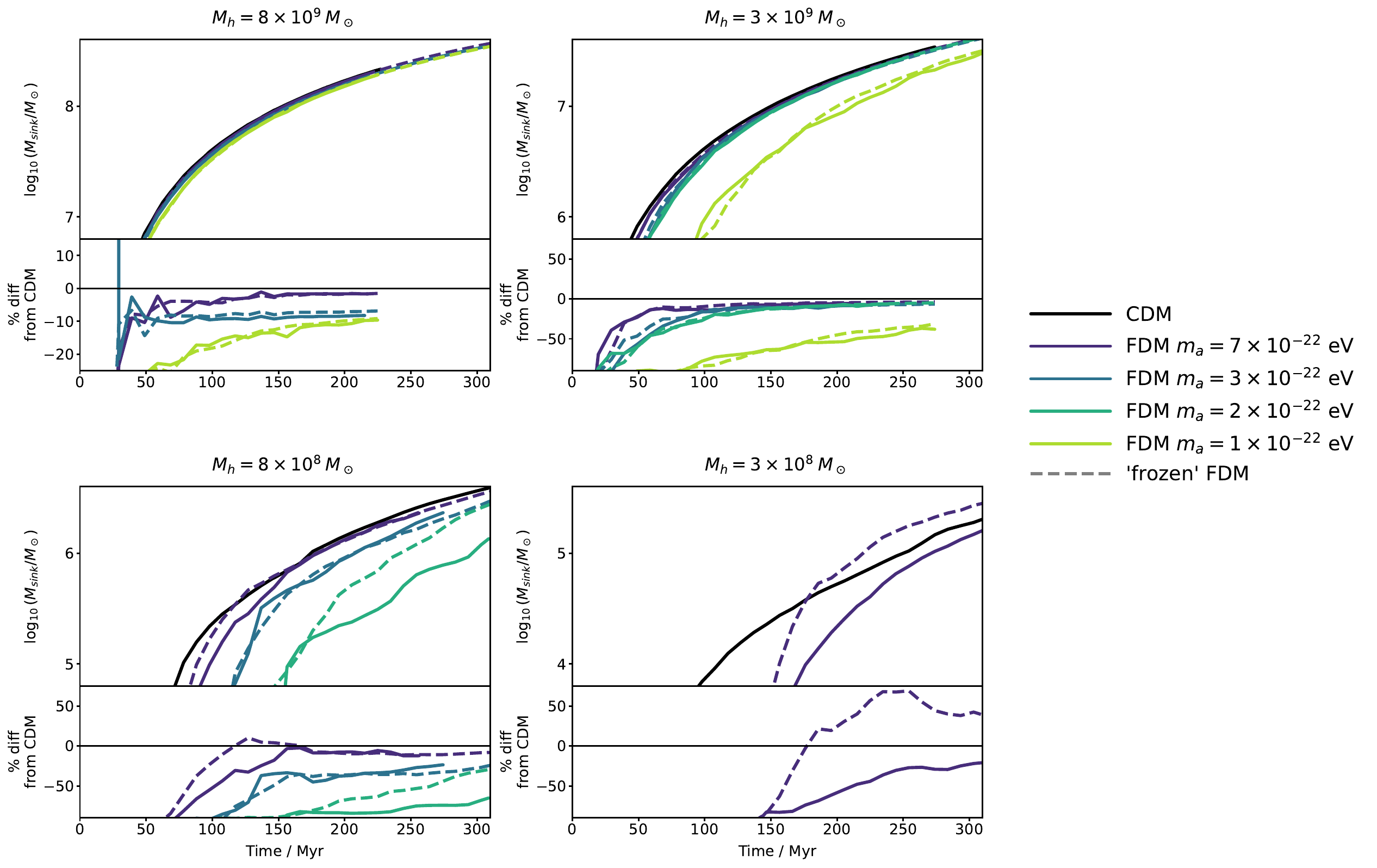}
    \caption{Cumulative mass in sinks over time across the four simulated halo masses (arranged in decreasing mass from top-left to bottom-right). Each of the four panels is divided into an upper plot showing the absolute total sink mass, and a corresponding lower plot displaying the percentage difference in mass accumulation relative to the CDM baseline. Dashed lines represent the \qmarks{frozen} FDM cases.}
    \label{fig:sink_mass}
\end{figure*}

Having looked qualitatively at how the gas is affected by the FDM dynamics, we now quantify the effects on the total accumulation of cold, dense gas. As detailed in \cref{sec:methods:code:sinks}, we employ sink particles as a numerical proxy for gas that has entered the runaway collapse phase ($n > \SI{e2}{\per\cm\cubed}$) representing pre-stellar collapse. This allows us to track the prospects of each halo to form stars and to make comparisons between haloes. \Cref{fig:sink_mass} illustrates the cumulative mass in sinks over time across our halo suite, providing a direct measure of the relative star-formation efficiency in FDM versus CDM environments.

\begin{table*}
    \centering
    \caption{Evolution of percentage reduction in accumulated sink mass relative to CDM. Percentages are calculated at three late-time snapshots ($t = \SIlist{200; 250; 300}{\Myr}$) after gas introduction. For our higher halo masses, the later times are not shown as the simulations did not run to that time due to computational time constraints. A value of \SI{-100}{\percent} indicates total suppression (no sinks formed), while positive values (e.\,g., the \SI{3e8}{\Msun}, $m_{\mathrm{a}} = \SI{7e-22}{\eV}$ \qmarks{frozen} case) indicate an increase in mass relative to CDM. Entries marked \qmarks{---} indicate that the simulation did not reach that time, or was not run entirely (as outlined in \cref{tab:simulation_parameters}).}
    \label{tab:suppression}
    \begin{tabular}{l S[table-format=3.0] *{4}{S[table-format=+3.0, retain-explicit-plus] S[table-format=+3.0, retain-explicit-plus]}}
    \toprule
    & & \multicolumn{8}{c}{\textbf{Mass Reduction relative to CDM (\%)}} \\
    {\textbf{Halo Mass}} & {\textbf{Time}} &
    \multicolumn{2}{c}{$m_{\mathrm{a}} = \SI{e-22}{\eV}$} & \multicolumn{2}{c}{$m_{\mathrm{a}} = \SI{2e-22}{\eV}$} & \multicolumn{2}{c}{$m_{\mathrm{a}} = \SI{3e-22}{\eV}$} & \multicolumn{2}{c}{$m_{\mathrm{a}} = \SI{7e-22}{\eV}$} \\
    {(\si{\Msun})} & {(\si{\Myr})} &
    {Dynamic} & {Frozen} & {Dynamic} & {Frozen} & {Dynamic} & {Frozen} & {Dynamic} & {Frozen} \\
    \midrule
    \multirow{1}{*}{$\mathbf{8 \times 10^9}$} & 200 & -11 & -10 & {---} & {---} & -8  & -7 & -2 & -2  \\
                                             
    \midrule
    \multirow{2}{*}{$\mathbf{3 \times 10^9}$} & 200 & -54 & -45 & -8 & -8  & -9 & -8 & -6 & -5  \\
                                              & 250 & -42 & -36 & -6 & -6  & -8 & -7 & -6 & -4  \\
    \midrule
    \multirow{3}{*}{$\mathbf{8 \times 10^8}$} & 200 & {---} & {---} & -84 & -68 & -37 & -36 & -8 & -9  \\
                                              & 250 & {---} & {---} & -76 & -54 & -28 & -35 & -12 & -11  \\
                                              & 300 & {---} & {---} & -70 & -32 & {---} & -28 & {---} & -9  \\
    \midrule
    \multirow{3}{*}{$\mathbf{3 \times 10^8}$} & 200 & {---} & {---} & {---} & {---} & -100 & -100 & -58 & +25  \\
                                              & 250 & {---} & {---} & {---} & {---} & -100 & -100 & -29 & +61  \\
                                              & 300 & {---} & {---} & {---} & {---} & -100 & -100 & -23 & +41  \\
    \bottomrule
    \end{tabular}
\end{table*}

\subsubsection{High-Mass Regime: Dominance of Halo Geometry}
In our most massive haloes ($M_{\mathrm{h}} = \SIlist{8e9; 3e9}{\Msun}$), the primary driver of suppression of gas in sinks is the cored geometry of the FDM potential. As shown in \cref{fig:sink_mass}, the mass accumulated in the \qmarks{frozen} FDM cases (dashed lines) closely tracks the fully \qmarks{dynamic} FDM runs (solid lines). 

For our heaviest halo \SI{8e9}{\Msun}, we observe a modest \SI{11}{\percent} reduction in total sink mass for the lightest simulated FDM case, $m_{\mathrm{a}} = \SI{e-22}{\eV}$, compared to CDM. In this regime, the gravitational potential is sufficiently deep that the stochastic fluctuations of the soliton act as a sub-dominant perturbation. Instead, the primary suppression comes from the flattened central potential gradient (the solitonic core), which reduces the gravitational acceleration in the central region, lowering the efficiency of gas inflow. After a brief initial delay in the onset of star formation, the sink accumulation rate reaches an approximately steady state relative to CDM, indicated by the flattening of the ratio in the lower sub-panels.

\subsubsection{Low-Mass Regime: Dominance of FDM Dynamics}
The impact of FDM physics becomes dramatically more pronounced in the lower-mass haloes ($M_{\mathrm{h}} = \SIlist{8e8; 3e8}{\Msun}$), where the dark matter dynamics emerge as the dominant suppression mechanism. 

For the $M_{\mathrm{h}} = \SI{8e8}{\Msun}$ case, we find up to an \SI{84}{\percent} reduction in sink mass for the \qmarks{dynamic} run compared to CDM obtained for the lowest axion mass we were able to simulate in this halo, $m_{\mathrm{a}} = \SI{2e-22}{\eV}$. Crucially, the \qmarks{frozen} FDM run for this same halo shows a significantly higher accumulation rate than its \qmarks{dynamic} counterpart. This gap confirms that core geometry alone cannot account for the suppression, and the time-varying dynamics of the fully-simulated FDM halo actively prevent gas from settling into the sink-forming phase.

In our lowest-mass haloes ($M_{\mathrm{h}} = \SI{3e8}{\Msun}$), star formation is almost entirely inhibited for lighter axion masses. Only the $m_{\mathrm{a}} = \SI{7e-22}{\eV}$ run formed sinks within the \SI{300}{\Myr} window, showing a substantial (\SI{23}{\percent}) but inconsistent reduction compared to CDM. For the other axion mass run that we performed for this halo mass ($m_{\mathrm{a}} = \SI{3e-22}{\eV}$), no sinks were formed within the first \SI{300}{\Myr} (the first sinks formed more than \SI{500}{\Myr} after the start of the simulation). This near-total suppression occurs because the de Broglie wavelength and hence soliton radius is comparable to the virial radius of the halo. In this state, the soliton fluctuations disrupt the entire central gas region, preventing the formation of a bound, cooling core. The percentage suppression in total sink mass for three late times (\SIlist{200; 250; 300}{\Myr} after the start of the simulation) are summarized for all cases in \cref{tab:suppression}.

Interestingly, in our lowest-mass halo ($M_{\mathrm{h}} = \SI{3e8}{\Msun}$), the \qmarks{frozen} FDM model ($m_{\mathrm{a}} = \SI{7e-22}{\eV}$) eventually surpasses the CDM case in total sink mass by nearly \SI{61}{\percent} at $t = \SI{250}{\Myr}$, despite experiencing a significant initial delay (see \cref{tab:sink_formation_delay}). This was the only FDM run in our suite to show such behaviour. We interpret this as the effect of a massive, static soliton core. In this lowest-mass regime, the soliton constitutes a dominant fraction of the total halo mass. When this mass is held static (as in the \qmarks{frozen} case), it creates a wide, deep, and perfectly stable potential well that, once gas cooling is initiated, facilitates a more efficient and monolithic accumulation of gas than the tidally-active CDM cusp. 

Crucially, this behaviour disappears entirely in the \qmarks{dynamic} FDM run, where the sink mass is suppressed by \SI{58}{\percent} at \SI{200}{\Myr} and remains significantly lower than CDM throughout. This stark contrast between the \qmarks{frozen} and \qmarks{dynamic} results at \SI{3e8}{\Msun} provides evidence that the suppression of star-forming gas in low-mass FDM haloes is not a geometric effect of the cored potential, but is driven by the kinetic energy and angular momentum injected by the stochastic soliton fluctuations.

\begin{table*}
    \centering
    \caption{Time delay of first sink formation ($t_\mathrm{delay}$) in FDM haloes relative to CDM. In CDM, sink formation starts at \SIlist{21; 12; 24; 60}{\Myr} (from the most massive halo to the lightest, in the order presented in the first row of the Table) from the moment the gas is introduced into the virialised halo. Cells marked \qmarks{---} indicate parameter combinations that were not simulated.}
    \label{tab:sink_formation_delay}
    \begin{tabular}{l *{4}{S[table-format=3.0] S[table-format=3.0]}}
    \toprule
    & \multicolumn{8}{c}{\textbf{Time for formation of first sink } $t_\mathrm{delay}$ (Myr)}  \\
    \textbf{Halo Mass} & \multicolumn{2}{c}{$m_{\mathrm{a}} = \SI{1e-22}{\eV}$} & \multicolumn{2}{c}{$m_{\mathrm{a}} = \SI{2e-22}{\eV}$} & \multicolumn{2}{c}{$m_{\mathrm{a}} = \SI{3e-22}{\eV}$} & \multicolumn{2}{c}{$m_{\mathrm{a}} = \SI{7e-22}{\eV}$}  \\
    (\si{\Msun}) & {Dynamic} & {Frozen} & {Dynamic} & {Frozen} & {Dynamic} & {Frozen} & {Dynamic} & {Frozen} \\
    \midrule
    $\mathbf{8 \times 10^9}$ & 0 & 1 & {---} & {---} & -2 & 0 & 1 & 1  \\
    \midrule
    $\mathbf{3 \times 10^9}$ & 18 & 50 & 6 & 10 & 13 & 6 & 6 & 7  \\
    \midrule
    $\mathbf{8 \times 10^8}$ & {---} & {---} & 89 & 63 & 55 & 59 & 37 & 31  \\
    \midrule
    $\mathbf{3 \times 10^8}$ & {---} & {---} & {---} & {---} & 631 & 473 & 79 & 75  \\
    \bottomrule
    \end{tabular}
\end{table*}

\subsubsection{Calibrating Semi-Analytical Models}

The observed flattening of the sink mass ratios at later times for our higher halo masses ($M_{\mathrm{h}} = \SIlist{8e9; 3e9}{\Msun}$) provides a vital calibration point for semi-analytical star formation models. Our results suggest that while star formation eventually occurs in FDM haloes, the efficiency is a steep function of both halo mass and axion mass, specifically at the low halo mass end of the FDM halo mass function of any FDM cosmology. Since the sink accumulation rate tracks the availability of cold, self-gravitating gas, we expect this suppression to translate into a proportionally reduced star formation rate. This implies that the transition from a dark, non-star-forming halo to a star-forming one is markedly different in FDM compared to CDM. We expect this change in star formation efficiency and distribution to significantly affect the predicted high-redshift UV luminosity functions and the timing and duration of reionization \citep{Barkana2001, Dayal2018, Mirocha2019}.

\subsection{Delay in First Sink Formation }
\label{sec:results:delay}

Our results indicate that the formation of the first sink is often delayed in FDM compared to CDM, even for haloes of the same mass, with the magnitude of the delay scaling inversely with both halo mass and axion mass. These delays in the formation of the first sink are shown in \cref{tab:sink_formation_delay}. In the high-mass regime ($M_{\mathrm{h}} = \SI{8e9}{\Msun}$), the delay is negligible ($|t_{\mathrm{delay}}| \lesssim \SI{2}{\Myr}$ across all models), as the deep gravitational potential quickly overcomes the pressure support of the gas even in the FDM haloes. 

However, in the low-mass regime ($M_{\mathrm{h}} \le \SI{8e8}{\Msun}$), the delay becomes much more significant across all the FDM cases when compared to the counterpart CDM halo. For these halo masses we see at least a \SI{30}{\Myr} gap between the first CDM sink formation and all of the FDM haloes. This delay is important to include when calibrating models of star formation in FDM cosmologies to the corresponding CDM cases, especially at high redshifts where a few tens of \si{\Myr} is a comparable timescale to the lifetimes of the first population of stars. A delay of $\sim \SI{100}{\Myr}$ in the emergence of the first stellar population would have a detectable impact on the timing of reionization \citep{Barkana2001, Robertson2015} and the structure of the 21-cm signal \citep[e.\,g.][]{Fialkov2014}.

By comparing the \qmarks{frozen} and \qmarks{dynamic} columns in \cref{tab:sink_formation_delay}, we can isolate the physical drivers of this delay. For $M_{\mathrm{h}} \gtrsim \SI{3e9}{\Msun}$, the delay times are similar for the \qmarks{frozen} and \qmarks{dynamic} cases, suggesting that the flattened potential gradient of the cored FDM halo profiles is the primary cause of the delay. In contrast, for our lighter haloes ($M_{\mathrm{h}} \le \SI{8e8}{\Msun}$) and the lightest axion masses, the \qmarks{dynamic} FDM run experiences a greater delay in first sink formation. For the lightest halo ($M_{\mathrm{h}} = \SI{3e8}{\Msun}$) and with $m_{\mathrm{a}} = \SI{3e-22}{\eV}$, the \qmarks{dynamic} FDM run is delayed by an additional $\sim \SI{160}{\Myr}$ relative to the \qmarks{frozen} case. This confirms that at small scales $r \lesssim r_{\mathrm{c}}$, the soliton fluctuations provide a secondary, dynamical barrier to collapse, likely through the injection of turbulent energy as hypothesized in \cref{sec:results:gas_distribution}.

To our knowledge, this is the first work to explicitly separate the effect of the geometric delay (core-driven) from the dynamical delay (fluctuation-driven) in a controlled simulation suite. While previous studies have noted a general delay in FDM star formation due to the power spectrum cut-off \citep[e.\,g.,][]{Hirano2018, Mocz2019, Jones2021}, our results confirm the importance of the dynamics at the halo centre, which, for the low $M_{\mathrm{h}}$ and $m_{\mathrm{a}}$ end of our parameter space, significantly extends the non-star-forming era of the smallest haloes beyond what a static cored-potential model would predict.

\subsection{Evolution of Sink Particles}
\label{sec:results:sinks}

\begin{figure*}
    \centering
    \includegraphics[width=\linewidth]{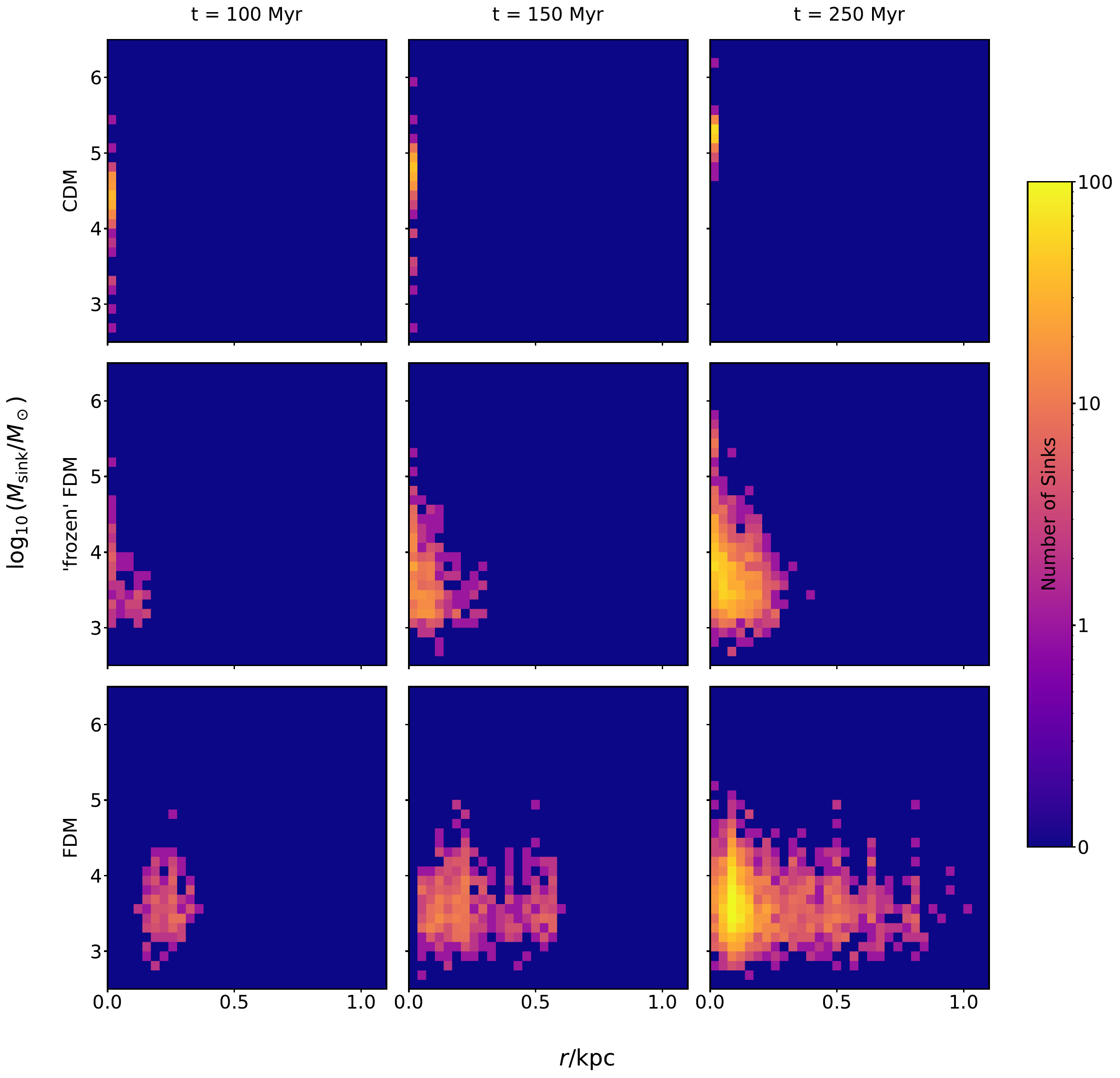}

    \caption{Time evolution of the sink particle population density in the mass–radius plane ($M_{\mathrm{sink}}$ vs.\ $r$) for a halo of mass $M_{\mathrm{h}} = \SI{3e9}{\Msun}$. Rows display the results for CDM (top), \qmarks{frozen} FDM with $m_{\mathrm{a}} = \SI{1e-22}{\eV}$ (middle), and the corresponding \qmarks{dynamic} FDM (bottom). The populations are shown at three characteristic epochs: shortly after first sink formation (\SI{100}{\Myr}), intermediate (\SI{150}{\Myr}), and late (\SI{250}{\Myr}). The colourbar indicates the number of sink particles per bin on a logarithmic scale. In the CDM case, the population remains a highly concentrated small number of sinks at the halo centre, showing a vertical growth track as sinks continue to accrete mass in the deep central potential. In contrast, the FDM cases show a distinct broadening of the radial distribution. While sinks in the \qmarks{frozen} potential tend to remain tightly around the static gravitational centre over time, the fully dynamical FDM case exhibits significant migration of sinks to larger radii.}
    \label{fig:sinks}
\end{figure*}

\begin{figure}
    \includegraphics[width=\linewidth]{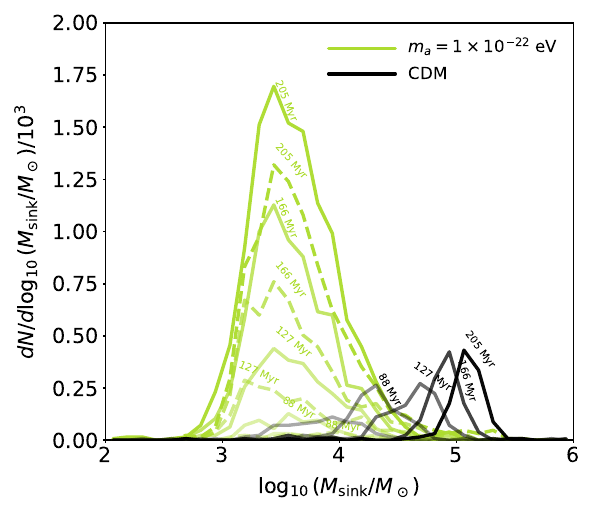}
    \caption{Evolution of the sink mass function for our \SI{3e9}{\Msun} halo, showing the FDM ($m_{\mathrm{a}} = \SI{1e-22}{\eV}$), \qmarks{frozen} FDM (dashed) and CDM (black) cases over time. It should be noted that the sinks do not represent individual stars, but cold dense gas with the potential to form stars. The sink mass function should therefore not be over-interpreted. The CDM case shows a rapid increase in the mass of each sink particle, with a much larger average sink mass by the end of the simulation, while the number of particles stays roughly the same. In contrast, in the \qmarks{dynamic} FDM and \qmarks{frozen} FDM cases, the number of sinks increases with time, but their masses do not evolve, resulting in an abundant population of low-mass sinks.}
    \label{fig:sinks_mass_function}
\end{figure}

Tracing the individual motion and growth of star-forming regions (using sink particles as a proxy) provides a direct probe of the small-scale dynamics within the dark matter potential. It is important to note that at our resolution, these sink particles represent dense, gravitationally bound gas clouds with the potential to form clusters of stars, rather than individual stellar objects. Consequently, the sink mass and the resulting mass function should be interpreted as indicators of the total reservoir of star-forming gas in a localized region rather than a direct prediction of the stellar Initial Mass Function (IMF). In this section, we examine the structural evolution of the sink population, focusing on the $M_{\mathrm{h}} = \SI{3e9}{\Msun}$ halo where the transition from core-dominated to dynamics-influenced behaviour is most notable. \Cref{fig:sinks} displays the number of sink particles in the mass–radius ($M_{\mathrm{sink}}$ vs.\ $r$) plane across three characteristic times: early (close to the first sink formation, \SI{100}{\Myr}), intermediate (\SI{150}{\Myr}), and late (\SI{250}{\Myr}).

\subsubsection{CDM: Centrally Concentrated Accretion}
In the CDM case (top row, \cref{fig:sinks}), the sink population follows the established picture of hierarchical collapse. Sinks typically form very close to the halo centre at $r < \SI{0.04}{\kpc}$ and remain there as the population of sinks have no means of gaining energy.

The vertical tracks in the $M_{\mathrm{sink}}$–$r$ plane indicate that CDM sinks are characterized by sustained accretion. As they settle into the stable potential minimum, they are continuously fed by accreting gas. This results in a \qmarks{top-heavy} sink mass function (\cref{fig:sinks_mass_function}), where the total number of sinks remains relatively stable increasing from $N = 133$ at \SI{100}{\Myr} to $N = 144$ at \SI{250}{\Myr}. However, the median sink mass grows significantly from $M_{\mathrm{med}} \approx \SI{3e4}{\Msun}$ at \SI{100}{\Myr} to \SI{2e5}{\Msun} at 250 Myr. This behaviour is consistent with the monolithic collapse characteristic of standard primordial star formation \citep[e.\,g.,][]{Bromm2013}.

\subsubsection{FDM: Dynamical Scattering and Starvation}
The evolution of the sink populations in FDM haloes is fundamentally different. While both the \qmarks{frozen} and \qmarks{dynamic} FDM cases show a delay in formation, their subsequent behaviours diverge based on the nature of the potential.

In the \qmarks{dynamic} FDM case (bottom row, \cref{fig:sinks}), we observe a significant migration of sinks to larger radii. Unlike the CDM sinks, which remain close to the halo centre, sinks in the \qmarks{dynamic} FDM haloes often move outwards to larger radii comparable to the core radius. As the soliton core undergoes its stochastic random walk, the site of the potential minimum shifts on timescales comparable to the gas cooling time. This leads to a population of sinks formed near the soliton radius ($r \approx r_{\mathrm{c}}$) that are subsequently ejected to larger radii by the time-varying gravitational potential. 

Crucially, once these sinks are scattered away from the dense central gas reservoir, their accretion ceases. This gas starvation results in a sink mass function that is markedly different from CDM: rather than a few massive sinks, the FDM halo hosts a growing number of low-mass clumps (increasing from $N = 154$ at \SI{100}{\Myr} to $N = 2801$ at \SI{250}{\Myr}). The median mass remains stagnant near the formation threshold ($M_{\mathrm{med}} \approx \SI{4e3}{\Msun}$), as can be seen by the lack of vertical growth in the bottom row of \cref{fig:sinks}, and in the much more bottom-heavy sink mass function in \cref{fig:sinks_mass_function}.

In the \qmarks{frozen} FDM case (middle row), we see an intermediate behaviour. While the sinks form at larger radii than in CDM due to the flattened core potential, they lack the extreme outward scattering seen in the \qmarks{dynamic} case. They tend to settle slowly toward the static gravitational centre, but their accretion is still less efficient than in CDM because the core's lower density reduces the gas supply rate.

This evidence highlights that small FDM haloes may harbour an altered population of the first stars. By scattering star-forming clumps out of high-density regions, FDM dynamics effectively limit the maximum mass any single star-forming region can attain, potentially shifting the initial mass function of Pop III stars toward lower masses.

The spatial distribution of these formation events and the subsequent growth history of individual sinks are further illustrated in \cref{app:sink_evolution}, where we show the sink population in radius–mass space. This visualization confirms that FDM sinks not only form at larger radii on average but see a general migration outwards in the halo.

\subsection{Chemical Stirring in the Central Halo Region}
\label{sec:results:H2}

While the initial collapse of gas into the virialised haloes in our mass range (\SIrange{e8}{e9}{\Msun}) is dominated by atomic hydrogen (Lyman-$\alpha$) cooling, the subsequent formation of stars is contingent upon the production of molecular hydrogen ($\mathrm{H}_2$). Efficient cooling to the temperatures needed for fragmentation ($T \lesssim \SI{500}{\K}$) requires the presence of $\mathrm{H}_2$ as the primary coolant \citep{Abel2002, Bromm2004, Glover2008, Bromm2011}. In this section, we examine how FDM dynamics alter the spatial and phase-space distribution of this crucial molecule, which has direct implications for the fragmentation and eventual IMF of the first stars.

\begin{figure*}
        \centering
        \includegraphics[width=0.9 \linewidth]{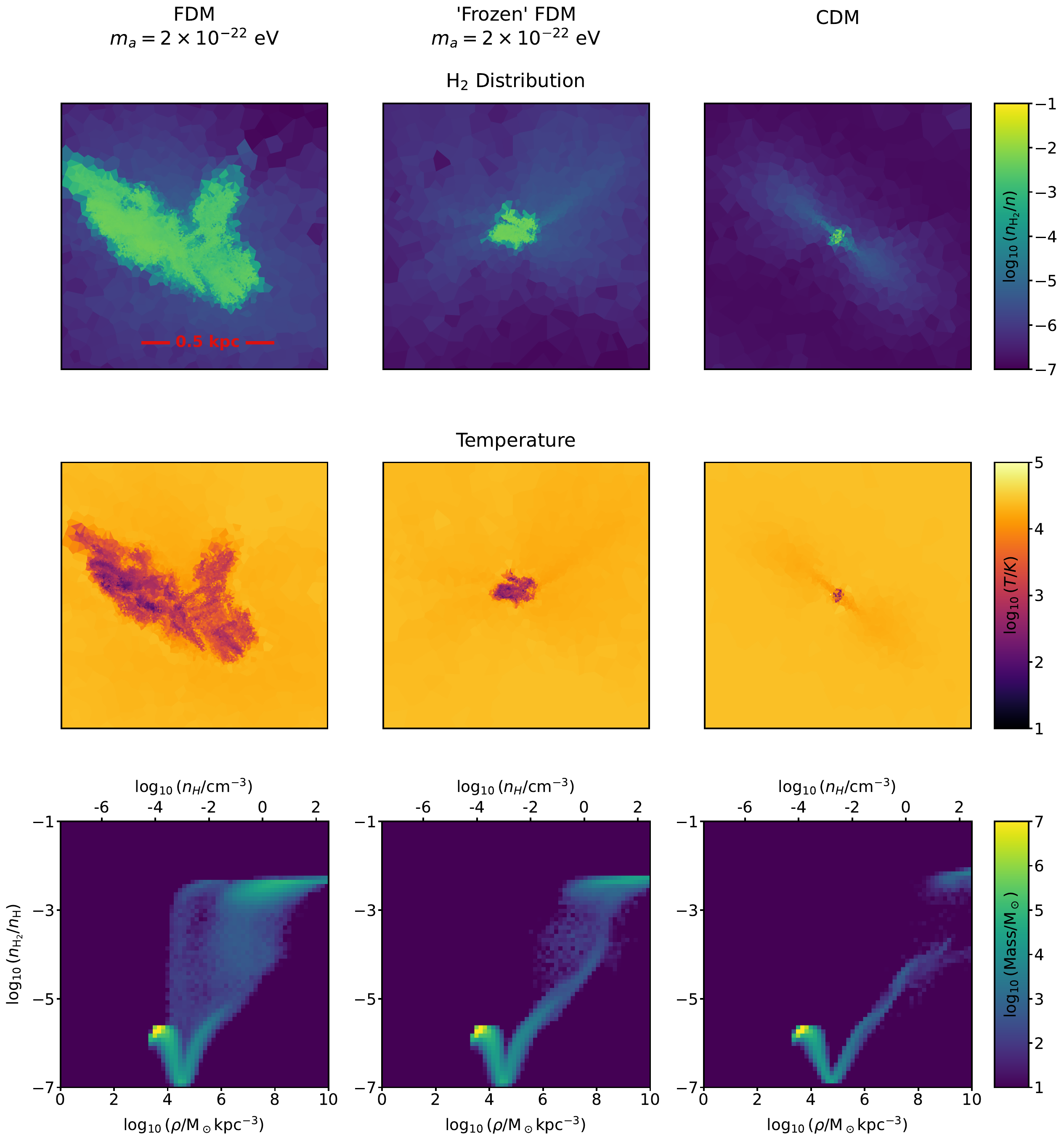}
        \caption{$\mathrm{H}_2$ properties at $t \approx t_* + \SI{190}{\Myr}$ (synchronized with the final row of \cref{fig:snapshot_series}), shown for a slice through a $M_{\mathrm{h}} = \SI{8e8}{\Msun}$ halo centre for FDM with $m_{\mathrm{a}} = \SI{2e-22}{\eV}$ (left), the corresponding \qmarks{frozen} FDM case (centre), and the comparison CDM case (right). We demonstrate the distribution of  $\mathrm{H}_2$ (top), the gas temperature of the same slice (middle), and the $\mathrm{H}_2$ concentration–density phase space (bottom). It is clear that in the FDM case, gas with a high concentration of $\mathrm{H}_2$ ($n_{\mathrm{H}_2}/n_{\mathrm{H}} > 10^{-3}$) is liberated much further from the halo centre than in either the \qmarks{frozen} FDM or the CDM case. This is evident from the bottom panels, where in the FDM case (left), concentrations of $n_\mathrm{\mathrm{H}_2}/n_\mathrm{H}\sim > 10^{-3}$ extends to much lower gas densities than either the \qmarks{frozen} FDM or CDM cases.}
        \label{fig:H2_distribution}
\end{figure*}

\Cref{fig:H2_distribution} shows the state of the gas for a $M_{\mathrm{h}} = \SI{8e8}{\Msun}$ halo, comparing the \qmarks{dynamic} FDM with $m_{\mathrm{a}} = \SI{2e-22}{\eV}$, the corresponding \qmarks{frozen} FDM, and CDM cases. We synchronize these snapshots with the final epoch of our density series (\cref{fig:snapshot_series}), though they represent a slice through the halo centre, rather than a projected density as in \cref{fig:snapshot_series}, to resolve the local chemical variations.

\subsubsection{FDM Stirring of H$_2$-enriched Gas}
In the \qmarks{dynamic} FDM case, $\mathrm{H}_2$-rich gas is liberated from the central potential. In standard CDM (right column), high $\mathrm{H}_2$ fractions ($n_{\mathrm{H}_2}/n_{\mathrm{H}} > 10^{-3}$) are strictly confined to the highest-density regions ($n \gtrsim \SI{10}{\per\cm\cubed}$) at the very centre of the halo. In contrast, the \qmarks{dynamic} FDM case exhibits plumes of $\mathrm{H}_2$-rich gas extending far into the lower-density outskirts of the halo, as can be seen in the top row of \cref{fig:H2_distribution}. 

We use the term \qmarks{liberated} to describe gas that was processed in the high-density core, where $\mathrm{H}_2$ formation is catalyzed, and subsequently transported to lower-density environments. This transport is driven by the stochastic \qmarks{stirring} of the solitonic core's random walk. As the potential minimum shifts, gas that was once at the centre is left in lower-density regions but retains its high $\mathrm{H}_2$ fraction. Because the cooling time is finite, this gas remains significantly colder than its surroundings, as evidenced by the middle row of \cref{fig:H2_distribution}, where the $T < 10^3$ K regions overlap almost perfectly with the redistributed $\mathrm{H}_2$.

This dynamical mixing is quantified in the bottom row of \cref{fig:H2_distribution}. In CDM, there is a clear, monotonic correlation: high $\mathrm{H}_2$ fractions only exist at high densities $\rho \gtrsim \SI{e9}{\Msun\per\kpc\cubed}$ ($n_{\mathrm{H}} \gtrsim \SI{10}{\per\cm\cubed}$). In the \qmarks{dynamic} FDM case, the distribution of high $\mathrm{H}_2$ concentration spreads much more broadly to the left, i.\,e.\ lower densities; high $\mathrm{H}_2$ concentrations ($n_{\mathrm{H}_2}/n_{\mathrm{H}} \sim 10^{-2}$) persist at densities as low as $\rho \sim \SI{e4}{\Msun\per\kpc\cubed}$ ($n_{\mathrm{H}} \sim \SI{e-3}{\per\cm\cubed}$).

The \qmarks{frozen} FDM case serves as a critical control. While it shows more extended $\mathrm{H}_2$ than CDM due to its cored potential, it lacks the broad, low-density \qmarks{tail} seen in the \qmarks{dynamic} run. This confirms that the $\mathrm{H}_2$ is not simply forming at larger radii due to the core shape, but is being actively moved by the time-varying gravitational potential.

\subsubsection{Implications for Fragmentation}

While we do not follow the gas down to higher densities in our simulations, this redistribution of cold gas has significant implications for the resulting stellar population. The thermodynamic state of primordial gas is a primary determinant of the fragmentation scale \citep{Bromm2004, Glover2013, KlessenGlover2023}. By spreading the $\mathrm{H}_2$ coolant over a wider volume, FDM dynamics create a more fragmented and turbulent environment. The presence of cold gas at lower densities suggests that the local Jeans mass, which scales as $M_{\mathrm{J}} \propto T^{3/2}\rho^{-1/2}$, is reduced in the outskirts compared to a centrally concentrated CDM model. This potentially favors the formation of a more \qmarks{bottom-heavy} IMF or a more spatially dispersed cluster of stars compared to the single, massive central stellar concentration predicted by standard CDM scenarios \citep{Stacy2011, Clark2011}. This aligns with our observation in \cref{sec:results:sinks} that FDM sinks are scattered and prevented from accreting into high-mass objects.

\section{Discussion}
\label{sec:discussion}

Our simulations reveal that the wave-like nature of FDM does not merely provide a static cored potential, but introduces a complex dynamical environment that fundamentally reshapes the birthplaces of the first stars. {While the monolithic $m_{\mathrm{a}} \sim 10^{-22}$~eV model is increasingly constrained by observations, our results establish a crucial first step to characterize the interplay between wave-driven dark matter fluctuations and baryonic dynamics and chemistry. By treating this mass range as a physical upper bound for wave-driven effects,} we have identified a new mechanism that suppresses gas accretion and redistributes key coolants. Below, we discuss the broader implications of these findings, {their relevance to more complex dark matter cosmologies,} and how these results compare to the wider cosmological landscape.

\subsection{Scope and Limitations}
\label{sec:discussion:scope}

Our simulations demonstrate that in isolated haloes, the inherent dynamical behaviour of FDM suppresses the accumulation of cold dense gas compared to CDM. These results represent a foundational step in understanding gas collapse in ultra-light dark matter haloes. The simplifying assumptions made here (specifically a spherically symmetric halo with gas collapsing from rest) do not capture the more complex, anisotropic structure formation reproduced by cosmological simulations, where gas possesses initial kinetic energy and turbulence, and collapsed structures are rarely self-contained or spherically symmetric \citep[e.\,g.][]{Mocz2019}. However, because the stochastic dynamics of the soliton are a universal feature of virialised FDM structures, we expect the qualitative conclusions regarding suppressed efficiency to be transferable to more comprehensively modeled environments.

This study targets the low-mass range of haloes, \SIrange{3e8}{8e9}{\Msun}, in which primordial stars form in FDM with $m_{\mathrm{a}} = \SIrange{e-22}{e-21}{\eV}$ considered here. However, these are still more massive than the $\sim \SI{e6}{\Msun}$ minihaloes expected for realistic primordial star formation in $\Lambda$CDM. Such haloes are heavily suppressed in FDM for the $m_{\mathrm{a}}$ range that we consider in this paper. We would need to consider heavier axions, if we were to simulate \SI{e6}{\Msun} haloes. However, such scenarios would require a different setup with much higher resolution, which would be computationally challenging in this setup, or a much smaller box size. Furthermore, these first star formation simulations must eventually include several feedback mechanisms unique to molecular cooling haloes, such as Lyman-Werner feedback \citep{Haiman1997}, the suppression of star formation by relative motion between dark matter and gas \citep{Schauer2021}, and internal turbulence \citep{Stacy2011}, which we leave for future work.

We have identified a region in the $M_{\mathrm{h}}$–$m_{\mathrm{a}}$ parameter space where FDM dynamics create signatures distinct from CDM. However, a finer sampling in $m_{\mathrm{a}}$, particularly in the regime $\SI{1e-22}{\eV} \lesssim m_{\mathrm{a}} \lesssim \SI{3e-22}{\eV}$ where the most interesting deviations from CDM were found, would allow for more reliable calibrations of star formation models. This change in gas distribution is especially evident in the transition from growing to shrinking sink half-mass radii observed in our simulations.

\subsection{Fragmentation, Chemical Stirring, and the Population III IMF}
\label{sec:discussion:chemistry_imf}

A unique finding of this work is the chemical stirring effect (\cref{sec:results:H2}), where FDM fluctuations transport $\mathrm{H}_2$-rich gas into low-density regions. While dark matter dynamics suppress the total gas accumulation, this redistribution suggests that star formation may be more spatially dispersed than in the monolithic CDM case. The presence of extra coolant far from the halo centre may have significant implications for the resulting stellar Population III IMF.

In CDM, the stable central potential favours the formation of a few massive sinks via sustained accretion. In contrast, the scattering and rotational support we observe in FDM results in a more fragmented, lower-mass sink population. It is reasonable to anticipate that such a significant modification to the sink population will be reflected in the characteristics and distribution of the first stars. 

While we can only speculate here about the effects of FDM dynamics on the true Pop~III IMF, increased levels of turbulence have been shown to promote fragmentation and lead to smaller stellar masses \citep{Clark2011, KlessenGlover2023}; the wave-driven \qmarks{stirring} in FDM may similarly shift the Pop~III IMF toward lower masses. Additionally, a broader region of gas enriched with $\mathrm{H}_2$ may further aid cooling and fragmentation with implications for the Pop~III IMF. However, further work is needed to follow the gas down to higher densities and smaller scales to isolate individual stars before robust conclusions can be made about any effect on the Pop III IMF.

\subsection{Comparison with Baryonic Feedback and Alternative Dark Matter}
\label{sec:discussion:comparison}

It is instructive to compare FDM dynamics with alternative solutions to the small-scale conflicts of $\Lambda$CDM. Recent literature \citep[e.\,g.,][]{Davies2020, Nori2024, Yang2024b} has explored how wave dark matter prevents gas/stars from settling, confirming the importance of the halo dynamics on the embedded baryons. In $\Lambda$CDM, baryonic feedback, such as supernova-driven potential oscillations, is often invoked to flatten central cusps \citep{Pontzen2012}. While feedback and inefficient star formation may explain why many low-mass subhaloes are dark \citep{Boylan2011}, it remains unclear if these solutions are universally effective in the lowest-mass galaxies \citep{Weinberg2015}.

Alternative models, such as Self-Interacting Dark Matter (SIDM), also address small-scale conflicts by creating flattened density cores through elastic particle scattering \citep{Tulin_Yu2018}. While both FDM and SIDM result in a suppression of the central gravitational cusp, they predict different relationships between halo mass and core properties, such as central density and core radius \citep{Mocz2023}. Since our results demonstrate that gas accumulation and fragmentation are highly sensitive to the specific geometry and depth of the central potential, it remains a compelling open question how the distinct core profiles of SIDM would affect the timing and scale of primordial star formation. A detailed comparison between the wave-driven dynamics of FDM and the scattering-induced profiles of SIDM would be essential for identifying unique baryonic signatures that could distinguish between these two dark matter candidates.

\subsection{Beyond Single-Field FDM}
\label{sec:discussion:sifdm}

The limitations of the single-field FDM model are becoming increasingly apparent. A fundamental challenge exists between the boson mass required to solve the small-scale problems of dwarf galaxies ($m_{a} \sim \SI{e-22}{\eV}$) and observational constraints, such as from dynamical heating of stars or from the Lyman-$\alpha$ forest, which favour a heavier boson mass \citep{Irsic2017, Marsh2019, Dalal2022, May2025}. This has motivated the exploration of more complex, multi-component or self-interacting dark matter. {These tight constraints on $m_a$ are predicated on the assumption that a single axion-like species constitutes \SI{100}{\percent} of the dark matter. MDM models, which combine CDM and FDM or multiple FDM fields of different masses \citep{Gosenca2023, Luu2024, Dome2025}, provide a promising pathway to reconcile these observations. As shown by \citet{Dome2025}, axion depletion in haloes below the FDM Jeans mass can effectively insulate dwarf galaxies from wave-driven heating, while still allowing for a globally significant $10^{-22}$~eV component. Our study therefore serves as the single-field limit for these more viable multi-component cosmologies.}

Similarly, self-interacting FDM (SIFDM) blends the wave-mechanical nature of axions with the collisional properties of self-interacting dark matter (SIDM), potentially alleviating the core-density tensions found in single-field models \citep{Mocz2023}. Scalar fields generically have quartic self-interactions that can modify the evolution of density perturbations, leading to a more effective flattening of central density profiles \citep{Mocz2023}. This represents a frontier of research that may provide a more complete picture of dark matter dynamics. We hope that the findings of this study can provide data and motivation for investigating, calibrating, and refining these next-generation dark matter simulations beyond the simple non-self-interacting, single-field FDM model.

\subsection{Implications for JWST and 21-cm Cosmology}
\label{sec:discussion:implications}

The suppressed star formation efficiency has direct consequences for large-scale observable signals. A delay or reduction in early star formation due to FDM dynamics naturally leads to a delay in the onset of both Cosmic Dawn and the Epoch of Reionization (EoR), thus shortening their duration relative to what is predicted in $\Lambda$CDM \citep{Nebrin2019, Jones2021, Liu2025, Dhandha2025b}. As a result, the cosmic thermal and ionization histories are altered, impacting the large-scale cosmological 21-cm signal of neutral hydrogen. Specifically, the modifications to the Population III IMF driven by FDM dynamics are expected to propagate into the X-ray heating of the intergalactic medium \citep{Sartorio2023}. This suggests a potentially stronger impact of FDM on the 21-cm signal than previously predicted, as the sensitivity of the 21-cm global signal and power spectrum to the Pop III IMF has been shown to be significant in the CDM context \citep{Gessey-Jones2025}. Consequently, future 21-cm experiments may provide robust avenues to constrain the axion mass to within \SI{10}{\percent} around \SI{e-21}{\eV} \citep{Liu2025}.

The recent discovery by JWST of an unexpectedly high number of massive, bright galaxies at $z > 10$ has sparked significant debate regarding the efficiency of early star formation. While the general delay in star formation found in FDM models might initially appear to exacerbate this tension, it is important to note that current observations primarily constrain relatively massive haloes ($M_{\mathrm{h}} > \SI{e9}{\Msun}$). At these scales, our simulations show a transition toward CDM-like behaviour, suggesting that the most luminous sources may be less affected by wave-like dynamics. However, the mass- and time-dependent star formation efficiency we have uncovered remains a critical variable; as suggested by \citet{Dhandha2025a}, a flexible efficiency that scales with halo mass can help reconcile theoretical predictions with JWST counts. In the low-mass regime where FDM dynamics are most potent, our predicted suppression remains an indicator for the faint-end of the UV luminosity function under FDM conditions.

\section{Conclusions}
\label{sec:conclusion}

In this study, we performed high-resolution hydrodynamical simulations of Fuzzy Dark Matter (FDM) using \textsc{axirepo} to investigate the impact of wave-like dark matter dynamics on the collapse and cooling of primordial gas. By comparing full dynamical simulations to \qmarks{frozen FDM} and CDM counterparts, we isolated the role of solitonic fluctuations in regulating star formation. Our simulations target haloes in the mass range \SIrange{3e8}{8e9}{\Msun}, a regime where the processes of halo formation, gas collapse, and star formation are strongly affected by FDM physics for axion masses of $m_{\mathrm{a}} = \SIrange{1e-22}{7e-22}{\eV}$. {We treat this range not as a claim for a monolithic dark matter candidate, but as a critical numerical benchmark to establish the maximal possible impact of wave dark matter in these haloes}.

We find that the stochastic random walk and mass fluctuations of the solitonic core act as a continuous source of gravitational perturbation. This leads to a substantial delay in the onset of first sink formation (the proxy for dense star-forming gas) and a significant reduction in total accumulated sink mass. While the delay in structure formation due to the FDM power spectrum cut-off is well-documented \citep{Schive2014, May2021, May2023}, we show for the first time through direct simulation that the internal dynamics of the halo provide a secondary, independent suppression mechanism that further delays and suppresses the onset of Cosmic Dawn. {By simulating the regime where FDM dynamics are most prominent, we establish an upper bound on the potential impact of FDM on the primordial environment. If we had found that the gas physics and star formation remain relatively unaffected in this extreme limiting case, it would provide a strong indication that such effects are negligible in more conservative or higher-mass models.}

Furthermore, our work reveals that FDM dynamics induce significant specific angular momentum in the gas within the soliton radius. For some parameter combinations at the low mass end of the $m_{\mathrm{a}}$ and $M_{\mathrm{h}}$ range explored, this acquired angular momentum reaches values comparable to the critical threshold ($j \gtrsim j_{\mathrm{crit}}$) required for rotational stabilization out to the soliton radius. This provides a physical barrier to collapse that is entirely absent in the CDM and \qmarks{frozen} FDM cases, demonstrating that soliton dynamics, rather than just the core's geometry, dictate the spatial scale of gas accumulation.

Our results demonstrate that the formation of the first sink is delayed in FDM compared to CDM counterparts, with the magnitude of this delay scaling inversely with both halo mass and axion mass. By comparing our \qmarks{frozen} and \qmarks{dynamic} models, we find that the flattened potential gradient of the cored FDM profile in the \qmarks{frozen} case is sufficient to explain this delay in most cases. However, the full wave-like fluctuations in the \qmarks{dynamic} models can cause an additional delay in the onset of star formation at the low-mass end of the $m_{\mathrm{a}}$ and $M_{\mathrm{h}}$ range examined in this work. In these cases, the FDM simulations experience a greater delay in first sink formation than their \qmarks{frozen} counterparts, confirming that soliton fluctuations provide a secondary, dynamical barrier that significantly delays gas collapse beyond what a static cored-potential would predict.

Within our simulations, we also see a dramatic shift in the distribution and growth of our sink particles, which in our simulations serve a proxies for dense star-forming regions. In CDM, sinks remain centrally concentrated and accrete mass efficiently. In our FDM haloes (both \qmarks{frozen} and \qmarks{dynamic}), sinks form at larger radii (near the soliton radius $r_{\mathrm{c}}$). However, in the \qmarks{dynamic} FDM case, these sinks are progressively scattered out to higher radii than in the \qmarks{frozen} comparison case.

We observe that FDM fluctuations extend the distribution of molecular hydrogen ($\mathrm{H}_2$) and cold gas into lower-density regions. This chemical stirring induces further cooling over a broader volume than is seen in CDM haloes. While the total gas collapse is suppressed, this dispersed cooling indicates a shift from monolithic central star formation to more fragmented, spatially diffuse clusters of dense, cold gas. This, coupled with our findings regarding the distribution and growth of sinks, hints that low-mass FDM haloes may naturally induce greater fragmentation and hence shift the Population III IMF toward lower masses compared to the massive stars predicted in CDM.

Our results provide an additional physical mechanism, beyond the large-scale power spectrum cut-off, leading to a delayed EoR in pure FDM cosmologies, {and motivates further study of these effects in more complex MDM models}. We suggest that the internal inefficiency of gas collapse, driven by both solitonic geometry and wave dynamics, further slows the contribution of the smallest FDM haloes to the ionizing budget. Critically, because this dynamical suppression is strongly mass-dependent and diminishes in more massive systems, our model suggests a high degree of variance in star formation efficiency across the halo mass function. This mass-dependent scaling offers a potential pathway to reconcile FDM with recent JWST observations of massive high-redshift galaxies \citep{Dhandha2025a}; while FDM strongly inhibits star formation in the smallest dwarf-precursors, it allows more massive haloes to assemble and form stars with efficiencies approaching those of $\Lambda$CDM, effectively shifting the balance of early light production toward the most massive systems.

Finally, our work demonstrates the necessity of incorporating accurate dark matter dynamics when deriving constraints from baryonic observables. This dynamical framework is essential for future investigations into multi-field or self-interacting FDM models, as we move toward a more complete description of dark matter {that better reflects the constraints on single-field FDM models}.

\section*{Acknowledgments}
AT acknowledges support from the UK Research and Innovation (UKRI) Science and Technology Facilities
Council (STFC) under Grant No.\ ST/W507362/1.
This work used the DiRAC Memory Intensive service Cosma8 at Durham University, managed by the Institute for Computational Cosmology on behalf of the STFC DiRAC HPC Facility (\texttt{www.dirac.ac.uk}). The DiRAC service at Durham was funded by BEIS, UKRI and STFC capital funding, Durham University and STFC operations grants. DiRAC is part of the UKRI Digital Research Infrastructure.
SCOG and RSK acknowledge financial support from the ERC via the Synergy Grant \qmarks{ECOGAL} (project ID 855130), from the German Excellence Strategy via the Heidelberg Cluster of Excellence (EXC 2181 - 390900948) \qmarks{STRUCTURES}. RSK furthermore thanks DFG and ANR for supporting project \qmarks{STARCLUSTERS} (ID KL 1358/22-1), and the German Ministry for Economy and Energy for funding project \qmarks{MAINN} (ID 50OO2206). 
The authors also thank Philip Mocz for his helpful comments on the paper.

%%%%%%%%%%%%%%%%%%%%%%%%%%%%%%%%%%%%%%%%%%%%%%%%%%
\section*{Data Availability}

Snapshot data and post-processed results are made available upon reasonable request.



\bibliographystyle{mnras}
\bibliography{fdm_star_formation} % if your bibtex file is called example.bib

%\begin{thebibliography}{99}
%\end{thebibliography}

%%%%%%%%%%%%%%%%%%%%%%%%%%%%%%%%%%%%%%%%%%%%%%%%%%

%%%%%%%%%%%%%%%%% APPENDICES %%%%%%%%%%%%%%%%%%%%%

\appendix

\section{Sink Formation and Evolution in Radius-Mass Space}
\label{app:sink_evolution}

\begin{figure}
    \centering
    \includegraphics[width=\columnwidth]{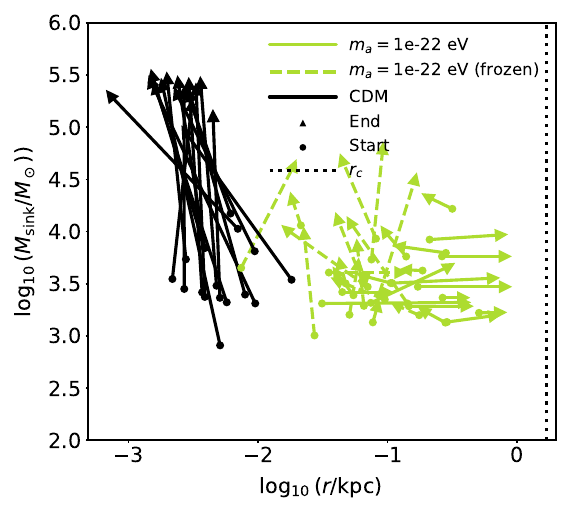}
    \caption{The distribution of sink particles in radius–mass space for representative CDM (black) and FDM (green) haloes. Circles denote the properties of sinks at the moment of formation, while the arrow ends indicate their final mass and radial position. Arrows indicate the evolutionary path of individual sinks, highlighting mass accretion and radial migration. In the CDM case, sinks quickly migrate toward the centre and grow into massive objects, whereas in the FDM cases, sinks remain more distributed with lower final masses, particularly in the \qmarks{dynamic} FDM (solid green) where a general migration outwards is seen}
    \label{fig:sink_paths}
\end{figure}

In this appendix, we provide a visualization of the sink particle lifecycle within our simulated haloes. \Cref{fig:sink_paths} shows the distribution of a random sample of 16 sink particles in the radius–mass plane, specifically tracking their initial formation (shown in the circles) and their final state at the end of the simulation (shown as the end of the arrows). These data were taken from our \SI{3e9}{\Msun} halo.

The figure highlights the distinct growth modes discussed in \cref{sec:results:sinks}. In CDM, sinks are primarily born near the centre of the deep gravitational cusp and rapidly accrete gas. In contrast, the FDM haloes show sink formation across a wider range of radii, coinciding with the interference-driven fluctuations of the solitonic core. The evolutionary tracks (arrows) in FDM show significantly less mass growth, and a general movement to larger radii (gravitational heating) confirming that wave-driven turbulence and angular momentum support limit the efficiency of gas accretion onto individual pre-stellar cores. In contrast, while the \qmarks{frozen} FDM is able to form sinks at larger radii than CDM, these sinks generally stay closer to the centre of the halo.

%%%%%%%%%%%%%%%%%%%%%%%%%%%%%%%%%%%%%%%%%%%%%%%%%%

% Don't change these lines
\bsp	% typesetting comment
\label{lastpage}
\end{document}